\documentclass[useAMS,usenatbib]{mn2e}
\usepackage{graphicx}
\usepackage{epsfig}
\usepackage{hyperref}
\newcommand{\beq}{\begin{equation}}
\newcommand{\eeq}{\end{equation}}
\newcommand{\beqa}{\begin{eqnarray}}
\newcommand{\eeqa}{\end{eqnarray}}

\def\eq#1{equation~(\ref{#1})}
\def\lexp{\mathop{{\langle}}\nolimits}
\def\rexp{\mathop{{\rangle}}\nolimits}

\def\bm{\bmath}

\def\r{{\bm{r}}}

\def\nbar{{\bar n}}
\def\Nbar{{\bar N}}
\def\cNbar{{\bar{\cal N}}}
\def\wbar{{\bar w}}
\def\Fbar{{\bar F}}
\def\xibar{{\bar \xi}}

\def\xbar{{\bar x}}

\def\d{{\rm d}}
\def\eff{{\rm eff}}

\def\min{{\rm min}}
\def\max{{\rm max}}

\def\deg{\,{\rm deg}}
\def\km{\, \textrm{km}}
\def\s{\, \textrm{s}}
\def\Mpc{\, \textrm{Mpc}}

\newcommand\fnurl[2]{%
  \href{#2}{#1}\footnote{\url{#2}}%
}

\begin{document}

\title[Hierarchical clustering in the CFHTLS wide]
{Evolution of hierarchical clustering in the CFHTLS-Wide since $z\sim1$\thanks{Based on observations obtained with
MegaPrime/MegaCam, a joint project of CFHT and CEA/IRFU, at the
Canada-France-Hawaii Telescope (CFHT) which is operated by the
National Research Council (NRC) of Canada, the Institut National des
Science de l'Univers of the Centre National de la Recherche
Scientifique (CNRS) of France, and the University of Hawaii. This work
is based in part on data products produced at Terapix available at the
Canadian Astronomy Data Centre as part of the Canada-France-Hawaii
Telescope Legacy Survey, a collaborative project of NRC and CNRS.}}

\author[M. Wolk et al.] {M. Wolk,$^1$\thanks{E-mail: wolk@iap.fr (MW)}
 H. J. McCracken,$^1$ S. Colombi,$^1$ J. N. Fry,$^{1,2} $ M. Kilbinger,$^3$ 
 \newauthor 
 P. Hudelot,$^1$ Y. Mellier$^1$ and O. Ilbert$^4$
\\
$^1$Institut d'Astrophysique de Paris, UMR 7095,
  98 bis Boulevard Arago, 75014 Paris, France\\ 
$^2$Department of Physics, University of Florida, Gainesville FL
 32611-8440, USA \\
$^3$CEA Saclay, Service d'Astrophysique (SAp), Orme des Merisiers,
 B\^{a}t. 709, 91191 Gif-sur-Yvette, France \\
$^4$Laboratoire d'Astrophysique de Marseille, UMR 7326, 
 38 rue Fr\'{e}d\'{e}ric Joliot-Curie, 13388 Marseille cedex 13, France}
\maketitle
\date{Released 2002 Xxxxx XX}

\begin{abstract} \\
We present measurements of higher order clustering of galaxies from
  the latest release of the Canada-France-Hawaii-Telescope Legacy
  Survey (CFHTLS) Wide. We construct a volume-limited sample of
  galaxies that contains more than one million galaxies in the
  redshift range $0.2<z<1$ distributed over the four independent
  fields of the CFHTLS. We use a counts in cells technique to measure
  the variance $\xibar_2$ and the hierarchical moments $S_{n}=
  {\xibar_n / \xibar_2^{n-1}}$ ($3 \leq n \leq 5$) as a function of redshift
  and angular scale.The robustness of our measurements if thoroughly tested, and the
field-to-field scatter is in very good agreement with analytical predictions.  At small scales,
  corresponding to the highly non-linear regime, we find a suggestion
  that the hierarchical moments increase with redshift.  At large scales, corresponding to the weakly non-linear
  regime, measurements are fully consistent with perturbation theory
  predictions for standard $\Lambda$CDM cosmology with a simple linear
  bias.
\end{abstract}


\section{Introduction}

In our current paradigm, structures in the Universe originate from tiny
density fluctuations emerging from a primordial Gaussian field after
an early inflationary period. Gravitational
instability in an homogeneous expanding Universe dominated by Cold
Dark Matter indeed gives rise to the rich variety of structures seen today by
hierarchical clustering.  This picture has shown to be
consistent with measurements of in million-galaxy surveys of the local
Universe like the Sloan Digital Sky Survey \fnurl{(SDSS)}{http://www.sdss.org/} and 2dF \citep{2dF}; however, our knowledge of the
evolution of the distribution of galaxies and its changing
relationship with underlying dark matter density field remains
incomplete.

On small scales, the evolution of the galaxy distribution strongly
depends on the physical processes of galaxy formation, while on larger
scales, both the initial power spectrum as well as to the global geometry
of the Universe play an important role. Precise measurements of the
galaxy distribution as a function of redshift and galaxy type
therefore represents a fundamental tool in cosmology and the $n$-point
correlation functions provide a particularly powerful method to
characterise it.

The two-point correlation function (along with is Fourier transform
the power spectrum) is the most widely used statistic
\citep[e.g.][for a general review on the subject]{Peebles80, Baumgart91, Martinez09}
as it provides the most basic measure of galaxy clustering -- the
excess in the number of pairs of galaxies compared to a random
distribution as function of angular scale. Nevertheless, the two-point
correlation function represents a full description only in the case of
a Gaussian distribution for which all higher-order connected moments
vanish by definition. It is well known that the \textit{actual} galaxy
distribution is non-Gaussian \citep[e.g.][]{FP78, Sharp84, Szs92,
  Bouchet93, Gaztanaga94}. Such a non-Gaussianity is already induced
by nonlinear gravitational amplification of mass fluctuations,  even
if they originated from an initial Gaussian field
\citep[e.g.][and references
therein]{Peebles80, Fry84, Juszkiewicz93, Bernardeauetal04}. But it
can also be present in initial conditions thereselves, from
topological defects such as textures or cosmic strings
\citep{Kaiser84} or fluctuations coming from certain inflationary
models \citep[][and references therein]{Guth81, Bartolo04}. In
addition, the process of galaxy formation results in a ``biased'' distribution of luminous matter with respect to
the underlying dark matter, which adds another level of complexity in
the non Gaussian nature of the process \citep[e.g.][and references
therein]{Politzer84, Bardeen86, Fryetal93, Mo96, Bernardeauetal04}. For all these reasons, high order moments
are required to obtain a full statistical description of the galaxy
distribution.

In this paper we will use the method of counts in cells to investigate
the evolution of the variance $\xibar_2$ and the hierarchical moments $S_{n}=
  {\xibar_n / \xibar_2^{n-1}}$ ($3 \leq n \leq 5$) of the galaxy distribution from $z \sim
1$ to present day. Until now, the majority of papers which have looked
at these quantities have analysed large local-Universe redshift
surveys; see \citet{RBM06,RBM07} for the SDSS and
\citet{Baughetal04,Crotonetal04,CNGB07} for the 2dFGRS. Furthermore,
only a few measurements have been made in the literature of the
redshift evolution of the hierarchical moments. For example,
\cite{Szapudietal01} used a purely magnitude-limited catalogue of
$\sim$ 710 000 galaxies with $I_{AB} < 24$ in a contiguous $ 4 \times
4 \deg^{2}$ region.  By assuming a redshift distribution for their
sources, they found that a simple model for the time evolution of the
bias was consistent with their measurements of the evolution of
$S_{3}$ as a function of redshift. They concluded that these
observations favoured models with Gaussian initial conditions and a
small amount of biasing, which increases slowly with redshift. Using
spectroscopic redshifts from the VVDS-Deep survey, \cite{Marinoni08}
reconstructed the three-dimensional fluctuation to $z \sim 1.5$ and
measured the evolution of $\bar{\xi}$ and $S_{3}$ over the redshift
redshift range $0.7<z<1$. They found that the redshift evolution in
this interval is consistent with predictions of first- and
second-order perturbation theory.

Measuring accurately the higher-order moments of the galaxy
distribution requires both very large numbers of objects and also a
precise control of systematic errors and reliable photometric
calibration: no previous surveys have the required combination of
depth and areal coverage to make reliable measurements from low
redshift to $z \sim 1$ in large volume-limited sample of galaxies.

In this work, we use data from the Canada-France-Hawaii Telescope Legacy Survey
(CFHTLS) which has a unique combination of depth, area and image
quality, in addition to an excellent and highly precise photometric
calibration. 

The paper is organized as follows. In Sect. 2, we present our data
set and the sample selection. In Sect. 3, we outline the techniques we
use to estimate the galaxy clustering starting from the two-point
correlation function and going to higher orders. The results are
described and interpreted in Sect. 4. Conclusions are
presented in Sect. 5. 

Throughout the paper we use a flat $\Lambda$CDM cosmology
($\Omega_{\rm m}=0.23$,  $\Omega_{\Lambda}=0.77$, 
$H_{0}= 100 h \km \s^{-1} \Mpc^{-1}$ and $\sigma_{8}=0.8$) 
with $h = 0.71 $. 
All magnitudes are given in the AB system.

\section{Observations, reductions and catalogue preparation}
\label{sec:observ-reduct-catal}

\subsection{The Canada-France-Hawaii Telescope Legacy Survey}

In this work, we use the seventh and final version of the
Canada-France-Hawaii-Telescope Legacy survey \fnurl{(CFHTLS)}
%Canada-France-Hawaii-Telescope Legacy survey}
{http://www.cfht.hawaii.edu/Science/CFHLS/}. 
The T0007 data release results in a major improvement of the absolute 
and internal photometric calibration of the CFHTLS.
Since T0007 is a reprocessing of previous releases, 
the actual area covered by this release is almost identical
to previous versions (some additional observations were added to fill
gaps in the survey created by malfunctioning detectors).

The improved photometric calibration has been achieved by applying
recipes adopted by the Supernova Legacy Survey
(SNLS)\fnurl{(SNLS)}{http://cfht.hawaii.edu/SNLS/} for the SNLS/Deep
fields to both the Deep and Wide fields of the CFHTLS \citep[for
details, see][]{Regnault:2009p12055}. This improved photometric
accuracy translates into improved photometric redshift precision and
in a consistent photometric redshift calibration over the four
disconnected patches of the CFHTLS Wide.

Complete documentation of the CFHTLS-T0007 release can be found at the
\fnurl{CFHT}{http://www.cfht.hawaii.edu/Science/CFHLS/T0007/} site.
In summary, the CFHTLS-Wide is a five-band survey of intermediate
depth.  It consists of 171 MegaCam deep pointings (of $ 1 \deg^2$
each) which, as a consequence of overlaps, consists of a total of $
\sim 155 \deg^2$ in four independent contiguous patches, reaching a
80\% completeness limit in AB of $u^*=25.2$, $g=25.5$, $r=25.0$,
$i=24.8$, $z=23.9$ for point sources.  The photometric catalogs we use
here were constructed by running
\fnurl{\texttt{SExtractor}}{http://www.astromatic.net/software/sextractor}
in ``dual-image'' mode on each CFHTLS tile, using the \textit{gri}
images to create a $\chi^2$-squared detection image
\citep[see][]{Szalay:1999p4804}. The final merged catalogue is
constructed by keeping objects with the best signal-to-noise from the
multiple objects in the overlapping regions. Objects are selected
using $i$-band \texttt{MAG\_AUTO} total magnitudes
\cite{1980ApJS...43..305K} and object colours are measured using $3$
arcsec aperture magnitudes.

After applying a magnitude cut at $i<22.5$, the entire sample contains
2,710,739 galaxies. After masking, the fields have an effective area of
\begin{equation}
S_{\eff, i} = 58.33, \quad  18.50, \quad 37.04, \quad 18.50 \deg^{2}, 
\label{eq:Seff}
\end{equation}
for fields W1, W2, W3, and W4 respectively, 
leading to a total area of $ 133 \deg^2$.

\subsection{Photometric redshift and absolute magnitude estimation}
\label{sec:phot-redsh}

The photometric redshifts we use here have been provided as part of
the CFHTLS-T0007 release. They are fully documented on the
corresponding pages at the
\fnurl{TERAPIX}{http://terapix.iap.fr/article.php?id_article=841} site
and the photometric redshift \fnurl{release documentation}
{ftp://ftpix.iap.fr/pub/CFHTLS-zphot-T0007/}
describes in details how the photometric redshifts were computed
following \citep{Ilbert:2006p709,Couponetal2009,Couponetal12}.

The photometric redshifts are computed using the \fnurl{``Le Phare''}
{http://www.cfht.hawaii.edu/~arnouts/lephare.html}
photometric redshift code which uses a standard template fitting
procedure. These templates are redshifted and integrated through the
transmission curves. 
The opacity of the intra-galactic medium (IGM) is accounted for and
internal extinction can be added as a free parameter to each
galaxy. The photometric redshifts are derived by comparing the
modelled fluxes and the observed fluxes with a $\chi^2$ merit
function. In addition, a probability distribution function is
associated to each photometric redshift.

The primary template sets used are the four observed spectra (Ell,
Sbc, Scd, Irr) from \citet{Colemanetal1980} complemented with two
observed starburst spectral energy distribution (SED) from
\citet{Kinneyetal1996}. These templates have been optimised using the
VVDS deep spectroscopic sample \citep{LeFevreetal05}. Next, an
automatic zero-point calibration has been carried out using
spectroscopic redshifts present W1, W3 and W4 fields.  For spectral
type later than Sbc, a reddening $E(B-V)=0$ to $0.35$ using the
\citet{Calzettietal2000} extinction law is applied.

Although photometric redshifts were estimated to $i<24$, we selected
objects brighter than $i=22.5$ to limit outliers \citep[see Table 3
of][]{Couponetal2009}. Star-galaxy separation is carried out using a
joint selection taking into account the compactness of the object and
the best-fitting templates.

Finally, using the photometric redshift, the associated best-fitting
template and the observed apparent magnitude in a given band, one can
directly measure the $k$-correction and the absolute magnitude in any
rest-frame band. Since at high redshifts the $k$-correction depends
strongly on the galaxy SED, it is the main source of systematic errors
in determining absolute magnitudes. To minimise the $k-$correction
uncertainties, we derive the rest-frame luminosity at a wavelength
$\lambda$ using the object's apparent magnitude closest to $\lambda
(1+z)$ according to the redshift of the galaxy \citep[the procedure is
described in Appendix A of][]{Ilbertetal05}.  For this reason the
bluest absolute magnitude selection takes full advantage of the
complete observed magnitude set.  However, as the $u$-band flux has
potentially larger photometric errors, we use $M_{g}$-band magnitudes.

\subsection{Sample selection}
\label{sec:sample-selection}

We select all galaxies with $i'<22.5$ outside masked regions. At
this magnitude, our photometric redshift samples are $100\%$ complete, 
for all galaxy spectral types, and our
photometric redshift errors are small compared to our redshift bin width in the redshift range
$0<z<1$.

Here we chose $M_{g}<-20.64$ ($M_{g}-5\log h<-19.9$) and divide this
selection into four redshift bins: $0.2<z<0.4$, $0.4<z<0.6$,
$0.6<z<0.8$ and $0.8<z<1.0$.
A large bin width ($\Delta z=0.2$) ensures a low bin-to-bin 
contamination from random errors, but still reduces substantially 
the mixing of physical scales at a given angular scale that occurs 
in an angular catalog without redshift information.

\begin{figure} 
\includegraphics[width=9cm]{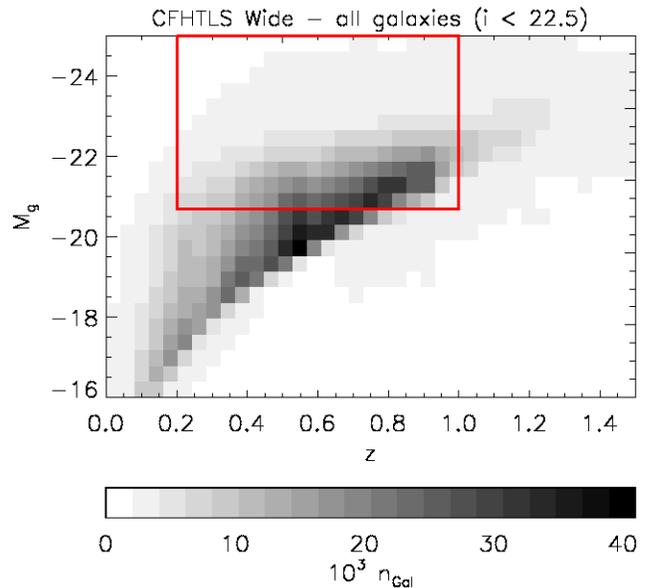} 
\caption{A two-dimensional histogram of the number density in the
  $M_{g}$-$z$ plane is plotted for all galaxies in the CFHTLS wide,
  T0007 release. The red line shows our adopted thresholds in absolute
  luminosity and redshift.}
\label{fig:selection}
\end{figure}

The sample selection is illustrated in Figure~\ref{fig:selection}.  As
a means of estimating the approximate completeness of our samples, we
measured the mean absolute magnitude $\bar{M}_{g}$ in each of the
redshift bins and found $\bar{M}_{g} = -21.4\pm 0.1$.  With this
selection we have more than 10,000 objects in each bin in each Wide
patch.  The detailed number of objects in each sample is summarized in
Table~\ref{table:objects}.

\begin{table}
\centering
\caption{Number of galaxies per redshift bin outside the masked regions.}
\begin{tabular}{ccccc}
\hline
     & W1 & W2 & W3 & W4 \\
\hline
$0.2<z<0.4$ & ~36 307 & 12 121 & ~23 426 & 10 945  \\
$0.4<z<0.6$ & 100 586 & 31 586 & ~67 413 & 32 778  \\
$0.6<z<0.8$ & 165 215 & 48 596 & 105 020 & 59 137  \\
$0.8<z<1.0$ & 149 960 & 45 783 & ~96 737 & 52 423  \\
\hline
\end{tabular}
\label{table:objects}
\end{table}

\section{Measuring galaxy clustering}
\label{sec:galaxy-clust-meas}

\subsection{Two-point angular correlation function}
\label{sec:two-point-angular}

As a consistency check with previous work, and as a means to verify
our lower-order counts in cells moments, we measure the two-point angular
correlation function $w(\theta)$ for our samples using the 
\citet{Landyetal93} estimator, 
\beq
w(\theta)=\frac{n_{\rm r}(n_{\rm r}-1)}{n_{\rm d}(n_{\rm d}-1)} 
 \frac{DD}{RR} - \frac{n_{\rm r}-1}{n_{\rm d}} \frac{DR}{RR} + 1,
\label{eq:LS}
\eeq where, for a chosen bin from $\theta$ to $\theta$ +
$\delta\theta$, $DD$ is the number of galaxy pairs of the catalog in
the bin, $RR$ the number of pairs of a random sample in the same bin,
and $DR$ the number of pairs in the bin between the catalog and the
random sample. $n_{d}$ and $n_{r}$ are the number of galaxies and
random objects respectively.

A random catalog is generated for each sample with the same geometry
as the data catalog using $n_{\rm r}=10^6$ which is at least 6 times
and at the most 90 times $n_{\rm d}$ for a given bin and patch. We
measure $w$ in each field using the \fnurl{``\texttt{Athena}'' tree
  code}{http://www2.iap.fr/users/kilbinge/athena/} in the angular
range $\theta \in [0.001,1]$ degrees divided into 15 bins spaced
logarithmically.

In \S~\ref{sec:comb-heir-momem} we explain how the individual
measurements of $w(\theta)$ from the four different fields are
combined.

\subsection{Higher-order moments}
\label{sec:Highorder}

Consider a distribution galaxy counts-in-cells which is a random
sampling of a continuous underlying density field $\rho(\mathbf{r})$
where $\mathbf{r}$ is the coordinate of the object.  The factorial
moments of the number of galaxies in a set of random cells of size
$\theta$ are closely related to the moments of the field through \beqa F_{k}
\equiv \lexp N^{[k]} \rexp = \lexp N (N-1) \cdots (N-k+1) \rexp
=\Nbar^k \bar{\rho}^{k} \eeqa where $\lexp N^{[k]} \rexp$ represents
the $k^{\rm th}$ factorial moment. $F_{1} = \lexp N
\rexp = \bar{N}$ represents the average cell count and $\bar{\rho} =
\rho \star W_{\theta}$ is the underlying density field smoothed with the window
function of the cells.

The quantities of interest, however, are the hierarchical amplitudes $S_{k}$ defined as 
\beq 
S_k = {\xibar_k \over \xibar_2^{k-1}}, 
\eeq 
where 
\beq
\bar{\xi}_{k}= \frac{1}{V^{k}}\int_{V}
 \d\r_{1} \dots \, \d\r_{k} \, \xi_{k}(\r_{1},\dots,\r_{k}) 
\eeq
is the $k$-point correlation function averaged over the cell angular
surface $V=s$ or volume $V=v$ according to whether statistics are considered 
in two or three dimensions, respectively.
In particular,
\beqa
% \bar{\xi}_{2}=\frac{1}{s^{2}}\int_{s}
% \int_{s}\,{\rm d} \mathbf{\theta}_{1}\,{\rm d} \mathbf{\theta}_{2}
% w(\bf{{\theta}}_{1},\mathbf{\theta}_{2}).
\bar{\xi}_{2}=\frac{1}{s^{2}} \int_{s} 
\d \theta_{1}\,\d \theta_{2} \, w(\theta_{1},\theta_{2}).
\label{eq:xiavdef}
\eeqa

The parameters $S_{3}$ and $S_{4}$, proportional to the skewness and
the kurtosis of the distribution of the numbers of galaxies in cells,
quantify deviations from the Gaussian limit. On a more general level,
the hierarchical amplitudes $S_{k}$ are expected to vary slowly with
angular scale as a consequence of gravitational instability both in
the weakly non-linear regime, as predicted by perturbation theory
\citep{Juszkiewicz93,Bernardeauetal04} and the highly non-linear
regime as shown by \cite{Davisetal77} by assuming stable clustering
plus self-similarity \citep[see also][and references
therein]{Peebles80,Bernardeauetal04}.

To relate hierarchical moments to factorial moments one can use 
the following recursion relation \citep{Szapudietal93, Colombietal00}: 
\beq
{S_{n}} = \frac{\xibar \, F_{n}}{N_{\rm c}^{n}} - \frac{1}{n} \, 
\sum_{k=1}^{n-1} \frac{n!}{(n-k)! \, k!} \, 
\frac{(n-k) \, S_{n-k} \, F_{k}}{N_{\rm c}^{k}}
\label{eq:Sn}
\eeq
with 
\beq
\xibar \equiv \bar{\xi}_{2} = {F_{2}}/{F_{1}^{2}} - 1
\label{eq:xibar}
\eeq
and $N_{\rm c} \equiv \bar{N} \bar{\xi}$.

In practice, one usually measures, for a cell of size $\theta$, the
count probability $P_{N}(\theta)$, the probability for a cell to
contain $N$ galaxies, as described below.  Once $P_{N}(\theta)$ is
measured, it is straightforward to calculate the factorial moments
using \beq F_k \equiv \sum_{N} P_{N}N^{[k]}. \label{eq:Fk} \eeq

\subsection{Measuring counts in cells}
\label{sec:BMW}

The main issue in counts in cells measurements is that large scales
are dominated by edge effects \citep{Szapudi96, Colombietal93} due to
the fact that, as cells have finite size, galaxies near the survey
edges or near a masked region have smaller statistical weights than
galaxies away from the edges. To correct for these defects, we use
``BMW-$P_N$'' \citep[Black Magic Weighted $P_N$;][]{ColombiSzapudi12,
  BSCBCDGPS06}.

This method is based on the fact that counts in cells do not depend
significantly on cell shape for a locally Poissonian point
distribution. This latter approximation allows one to distort the
cells near the edges of the catalog and near the masks in order to
reduce as much as possible edge effects discussed above. In practice,
the data and the masks are pixellated on a very fine grid. In this
case, one can consider all the possible square cells of all possible
sizes. For each of these cells, an effective size is given which
corresponds to the area that is contained in the cell after
subtraction of the masked pixels. If the fraction of masked pixels in
the cell exceeds some threshold $f$, the cell is discarded. Then, the
center of mass for the unmasked part of the cell is calculated and the
algorithm finds the corresponding pixel it falls into.  As one cell
per pixel is enough to extract all the statistical information, the
code selects among all the cells of equal unmasked volume targeting
the same pixel the most compact one, i.e. the one which has been the
least masked.

Note that the current implementation of BMW-$P_N$ works only in the small
angle approximation regime. This is not an issue in our analyses as
the largest field W1 has an angular extension smaller than 9 degrees.

We used the following parameters in BMW-$P_N$: the sampling grid
pixel size was set to 1.93 arcsec, which corresponds to a grid of size
$16000 \times 13944$ for W1. This very high resolution allowed us to
probe in detail the tails of the count in cells probability at
scales as small as $\theta_\min=0.0011 \deg $.  We measured counts in
square cells of size $\theta$ for 15 angular scales, logarithmically
spaced in the range $\theta \in [\theta_{\min}, \theta_{\max}]$ with
$\theta_{\rm max}=1 \deg $ (see Tables \ref{table:Sns}).
 
The allowed fraction of masked pixels per cell $f$ is a crucial
parameter. If $f$ is too large, the cells become too elongated,
invalidating the local Poisson approximation. On other hand, if $f$ is
too small, edge effects become prominent.  We investigated the
following set of values of $f$ for W1 in the redshift bin $0.6<z<0.8$:
$f=0$, which corresponds to a traditional counts in cells measurement
with square cells not overlapping with the edges of the catalog or
the masks, and $f=0.25,0.5,0.75$. We found that $f$ as large as 0.5 is
needed to ensure that edge effects are minimised in the scaling range
probed by our measurements. Larger values of $f$ do not change the
results significantly so we kept $f=0.5$ to reduce as much as possible
cell shape anisotropies.

\subsection{Combining fields}
\label{sec:comb-heir-momem}

Factorial moments obtained from equation (\ref{eq:Fk}), where counts
in cells are measured on a single field W${i}$, are unbiased estimators. 
However, concerning counts in cells, the quantities of interest 
are the averaged two-point correlation function $\xibar$ and the 
hierarchical moments $S_{n}$, which are non linear combinations of $F_{n}$, 
as in equations (\ref{eq:Sn}) and (\ref{eq:xibar}), and hence biased 
statistics \citep[e.g.][]{Szapudi99}.
To minimise the bias, it is preferable to first perform a weighted average 
of the factorial moments over the four fields with weights proportional 
to effective area $ S_{\eff,i} $, 
\beq
\Fbar_{n} = \frac{\sum_{i} S_{\eff,i} F_{n,i}}{\sum_{i} S_{\eff,i}}, 
\label{eq:combined}
\eeq
where $F_{n,i}$ is the $n^{\rm th}$ factorial moment measured on
field $i$ and  $S_{\eff,i}$ its effective area 
(equation \ref{eq:Seff}) --
and then, to use equations (\ref{eq:Sn}) and (\ref{eq:xibar}) to
compute $\xibar$ and $S_{n}$ from $\Fbar_{n}$. 

We have decided to use the same weighting procedure as in equation 
(\ref{eq:combined}) to combine the individual correlation functions 
$w_{i}(\theta)$ measured in each field and derive a combined $w(\theta)$.
This simpler approach leads to slightly biased measurements but is 
sufficient for our purpose. 

To calculate error bars, we exploit the fact that we have at our disposal 
four independent fields to estimate statistical uncertainties 
directly from the data, 
using the dispersion among the four fields.
The calculation of such errors is potentially complicated, as each 
field has a different area.
For the factorial moments, which are unbiased, the errors are 
\beq
(\Delta F_{k})^{2} 
= \frac{1}{(n_{\rm fields}-1)}
\sum_{i} S_{\eff,i} (F_{k,i}-\Fbar_{k})^{2} / \sum_{i} S_{\eff,i} .  
\label{eq:errFn}
\eeq
Details are provided in Appendix \ref{app:stat}. 

Rigorously speaking, a full error propagation formula should 
be used to compute the statistical uncertainties of $w(\theta)$ 
\citep{Landyetal93, Bernstein94}, 
and of $\xibar$ and $S_{n}$ \citep{Szapudi99}.
However in our case an involved procedure such as this is unnecessary
since the uncertainties on the estimated errors are very large due to
the fact that we have access to only four independent fields.  To keep
the approach simple while preserving, in practice, sufficient
accuracy, we compute the errors on $w(\theta)$, $\xibar$ and $S_{n}$
as for the $ F_k $, by simply replacing $F_{k,i}$ with $w_{i}(\theta)$,
$\bar{\xi}_{i}$ or $S_{k,i}$ in equation (\ref{eq:errFn}) ($i$ is
the field number), where the quantities $\xibar_{i}$ and $S_{k,i}$ are
derived from $F_{k,i}$ using equations (\ref{eq:Sn}) and
(\ref{eq:xibar}).

The hierarchical moments and their associated error bars are presented
for each redshift bin in Tables \ref{table:Sns}.

\subsection{Transformation to physical co-ordinates}
\label{sec:projection}
Transformation to physical co-ordinates is treated following \citet{GP77} 
and \citet{FP78}, originally due to \citet{L53}.
Let $ \nbar $ be the spatial density of galaxies and 
$ \phi $ be the selection function, that is, the probability 
that a galaxy at comoving distance $x$ or redshift $z$ 
is included in the catalog.
Summed over distances in a spatially flat universe, 
the projected number density $ \cNbar $ per solid angle 
in a given redshift range $ z_\min < z < z_\max $ is 
\beq
\cNbar = \int a^3 x^2 {\rm d}x \, \nbar \phi = \int_{z_\min}^{z_\max} n(z) \, {\rm d}z , 
\eeq
where a photon at redshift $z$ 
arrives from comoving distance 
\beq
x(z) = c H_0^{-1} 
\int_0^{z} {{\rm d}z' \over \sqrt{ \Omega_m (1+z')^3 + \Omega_\Lambda}} .
\eeq
The density of galaxies per solid angle per redshift interval is 
\beq
n(z) = {{\rm d}N \over {\rm d}z \, {\rm d}\Omega} = \nbar a^3 x^2 \,
{{\rm d}x \over {\rm d}z} \, \phi.
\eeq
The factor $ \nbar a^3 $ is constant, and
we take $ a(z=0) = 1 $.
Each sample has a characteristic depth weighted by 
geometry and redshift distribution, 
\beq
x^* = {\int n(z) \, x(z) \, {\rm d}z \over \int n(z) \, {\rm d}z} , 
\eeq
where the integrals run from $z_\min$ to $z_\max$.
The angular density scales as $ \cNbar \sim (x^*)^3 $.

The projected two-point function $ w(\theta) $ is related 
to the spatial two-point function $\xi(x) $ as 
\beq
\cNbar^2 w(\theta) = \int n(z_1) \, {\rm d}z_1 \, n(z_2) \, {\rm d}z_2 \; 
\xi(x_{12}) . \label{w2}
\eeq
The correlation function is important only at small separations, 
for which $ x_{12} $ is 
\beq
x_{12}^2 = \bigl( {{\rm d}x \over {\rm d}z} \bigr)^2 \, \Delta z^2 
 + x^2 \, \Delta \theta^2 . 
\eeq
Changing variables to  $ z = \frac12 (z_1+z_2) $ and 
$ \Delta z = z_1-z_2 $, and defining $y$ as 
$ y = ({\rm d}x/{\rm d}z) \Delta z /x \theta  $ leads to
\beq
\cNbar^2 w(\theta) = \int {\rm d}z \, n^2(z) \, \int 
{x \, \theta \, {\rm d}y \over {\rm d}x/{\rm d}z} \; 
\xi\bigl( x \theta (1+y^2)^{1/2} \bigr) .
\eeq
To the extent that $x^*$ is the characteristic scale for $x$, 
the second moment scales as 
$ w(\theta) = x^{*-1} \, {\rm Fct} (\theta x^*) $; that is, 
the combination $ x^* w $ is a function of $ \theta x^* $ 
\citep[cf.][]{GP77}.

The projection depends on the form of the spatial correlation function.
The observed correlation function is approximately a power law; 
including the stable clustering time dependence, 
$ \xi = a^{3-\gamma}  (x/x_0)^{-\gamma} $, 
we have 
\beq 
w(\theta) = C_\gamma \, \theta^{1-\gamma} \, x_0^\gamma \; 
{\int {\rm d}z \, n(z)^2 \, a^{3-\gamma} \, x^{1-\gamma} \, ({\rm
    d}x/{\rm d}z)^{-1} 
 \over \left[ \int {\rm d}z \, n(z) \right]^2 } , 
\eeq
where 
\beq
C_\gamma = \int_{-\infty}^\infty {{\rm d}y \over (1+y^2)^{\gamma/2} }
= {\Gamma({1 \over 2}) \, \Gamma({\gamma - 1 \over 2}) \over 
\Gamma({\gamma \over 2}) } \approx 3.67909 , 
\eeq
approximated for $ \gamma \approx 1.8 $.
Thus, the angular two-point function is also a power law, 
\beq
w(\theta) = {I_2 \over I_1^2} \, x_0^\gamma \, \theta^{1-\gamma} , 
\eeq
with power index shifted by 1.

The projected $n$-point function is given by the extension of \eq{w2}, 
\beqa
\cNbar^k w_n  &=& \int n_1 \, {\rm d}z_1 \dots n_k \, {\rm d}z_k \, \xi_n .
\eeqa
For hierarchical model where the $n$-point correlation function can be
expressed as a sum over products of 2-point correlation functions: $ \xi_n = \sum Q_k \, \xi_{ij}^{k-1} $ 
(sum over connected terms), where $Q_k$ are the reduced amplitudes and
$\xi_{ij} \equiv \xi_{2}(\bf{x}_{i}, \bf{x}_{j})$.
After a change of variables to 
the average redshift and $ k-1 $ differences, 
the angular correlation function is also found to be hierarchical, 
with amplitude $ q_k = Q_k F_{\rm p}(k) $, 
\beq
F_{\rm p}(k) = {I_k \, I_1^{k-2} \over I_2^{k-1} } , \label{Fpn}
\eeq
\beq
I_k = C_\gamma^{k-1} \int n^k \, {\rm d}z \, 
\Bigl( {a^{3-\gamma} x^{1-\gamma} \over {\rm d}x/{\rm d}z} \Bigr)^{k-1} \label{In} .
\eeq
Table \ref{table:Fpn} shows numerical results for the redshift 
distribution of the data with $ \gamma = 1.8 $.
Variations from field to field are generally less than a percent.
Changing $\gamma$ by 10\% affects the results by at most a few percents,
and considering 
the perturbation theory time dependence $ \xi(x) \sim D^2(z) $
instead of $ a^{3-\gamma} $ by less than 1\%.

\begin{table}
\centering
\caption{
Projection parameters. Columns give the redshift range, 
scaling distance
and $n$-point amplitude projection factors 
$F_{\rm p}(n) $ (eq.~[\ref{Fpn}]) for $ n = 3 $, 4, 5.}
\begin{tabular}{ccccccc}
\hline
$z_1$ & $z_2$ & $x^*$ (Mpc) & 
 $F_{\rm p}(3)$ & $F_{\rm p}(4)$ & $F_{\rm p}(5)$ \\
\hline
0.2 & 0.4 & $ 1280 $ & 1.09 & 1.28 & 1.61 \\
0.4 & 0.6 & $ 1910 $ & 1.02 & 1.06 & 1.12 \\
0.6 & 0.8 & $ 2530 $ & 1.02 & 1.05 & 1.10 \\
0.8 & 1.0 & $ 3010 $ & 1.09 & 1.25 & 1.46 \\
\hline
\end{tabular}
\label{table:Fpn}
\end{table}

\subsection{Effect of photometric errors}

In order to quantify the effect of the photometric errors on the
values of the $S_{n}$, we investigate how important the variation is
in the projection factors.  To do so we use results from
\cite{Couponetal2009} who compute the photometric redshifts for the CFHTLS Wide
T0004. They use a template-fitting method to compute photometric
redshifts calibrated with a large catalog of 16,983 spectroscopic
redshifts from the VVDS-F02, VVDS-F22, DEEP2 and the zCOSMOS
surveys. Their method includes correction of systematic offsets,
template adaptation, and the use of priors. Comparing with
spectroscopic redshifts, they find a photometric dispersion in the
wide fields of 0.037-0.039 and at $i'<22.5$. We then take a dispersion
$\Delta z = 0.038 \, (1+z)$, and we get the numbers presented in Table
3 for x*, $Fp_2$, $Fp_3$, $Fp_4$, $Fp_5$.  We see that the errors on
the photometric redshifts introduce a very small difference in the
projected numbers and we can consider that they are negligible
comparing to the error that we get from the field
to field variance.\\

\begin{table*}
%\begin{minipage}{126mm} 
\centering
\begin{tabular}{cccccccc}
\hline
$z_1$ & $z_2$ & $x^*$ (Mpc) & 
 $F_{\rm p}(3)$ & $F_{\rm p}(4)$ & $F_{\rm p}(5)$ \\
\hline
0.2 & 0.4 & $ 1272 $ & 1.09 & 1.29 & 1.64 \\
0.4 & 0.6 & $ 1905 $ & 1.02 & 1.07 & 1.14 \\
0.6 & 0.8 & $ 2530 $ & 1.02 & 1.06 & 1.11 \\
0.8 & 1.0 & $ 3005 $ & 1.09 & 1.25 & 1.48 \\
\hline
\end{tabular}
\label{table:Fndelta}
\caption{Projection parameters obtained using photometric redshifts 
and the dispersion in the estimated redshift of 
$\Delta z = 0.038 \, (1+z)$. Columns give the redshift range, 
scaling distance
and $n$-point amplitude projection factors $F_{\rm p}(n) $ (eq.~[\ref{Fpn}]) 
for $ n = 3 $, 4, 5.}
\end{table*}

\section{Results}
\subsection{Two-point angular correlation function}

\begin{figure}
\includegraphics[width=8cm]{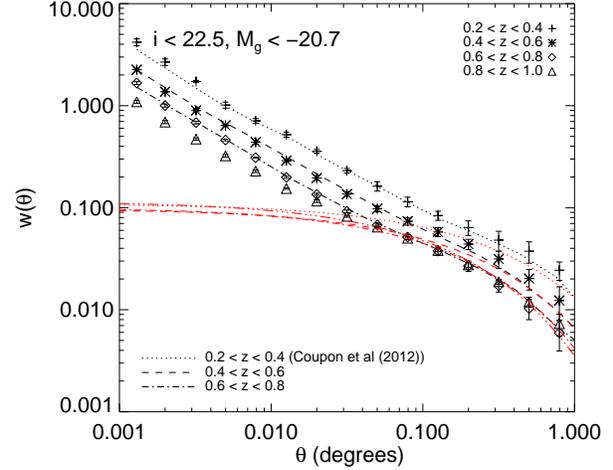}
\caption{The two point angular correlation function, $w(\theta)$
  (symbols) compared with the measurement of Coupon et al. (dotted,
  dashed and dot-dashed black curves). The red curves represents the
  linear theory prediction taking into account linear bias, as
  discussed in Section~\ref{sec:comb-meas}.}
\label{figure:wcomb}
\end{figure}

As a first consistency check, Figure~\ref{figure:wcomb} shows our
measurements of the two-point correlation function $w(\theta)$ in four
redshift bins. We compare three bins with those of \cite{Couponetal12}
(dashed lines). On the scales shown in this Figure, agreement between
the two measurements is excellent. This is to be expected: at $i<22.5$
the precision of photometric redshifts in CFHTLS release T0007 is
similar to the previous release, T0006, used in
\cite{Couponetal12}. We over-plotted the dark matter
predictions for the linear part of the two-point correlation function
from which we determine the value of the bias in the different
redshift bins. These are presented in Table 4. One can note that our
values for the bias are close to unity do not strongly evolve with
time.

\begin{figure*}
  \begin{center}
    \begin{tabular}{c@{}c@{}}
      \includegraphics[width=8cm]{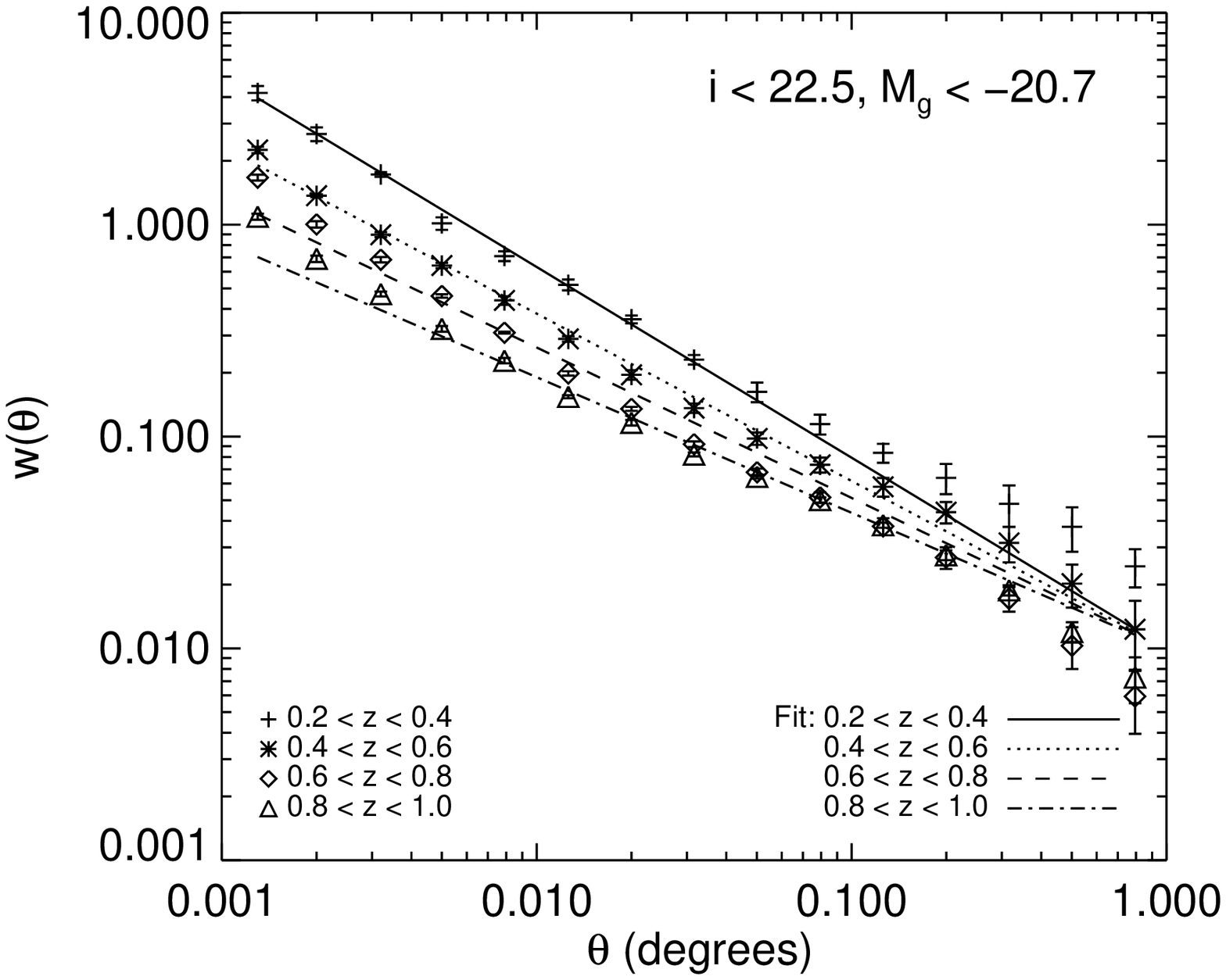}
      \includegraphics[width=8cm]{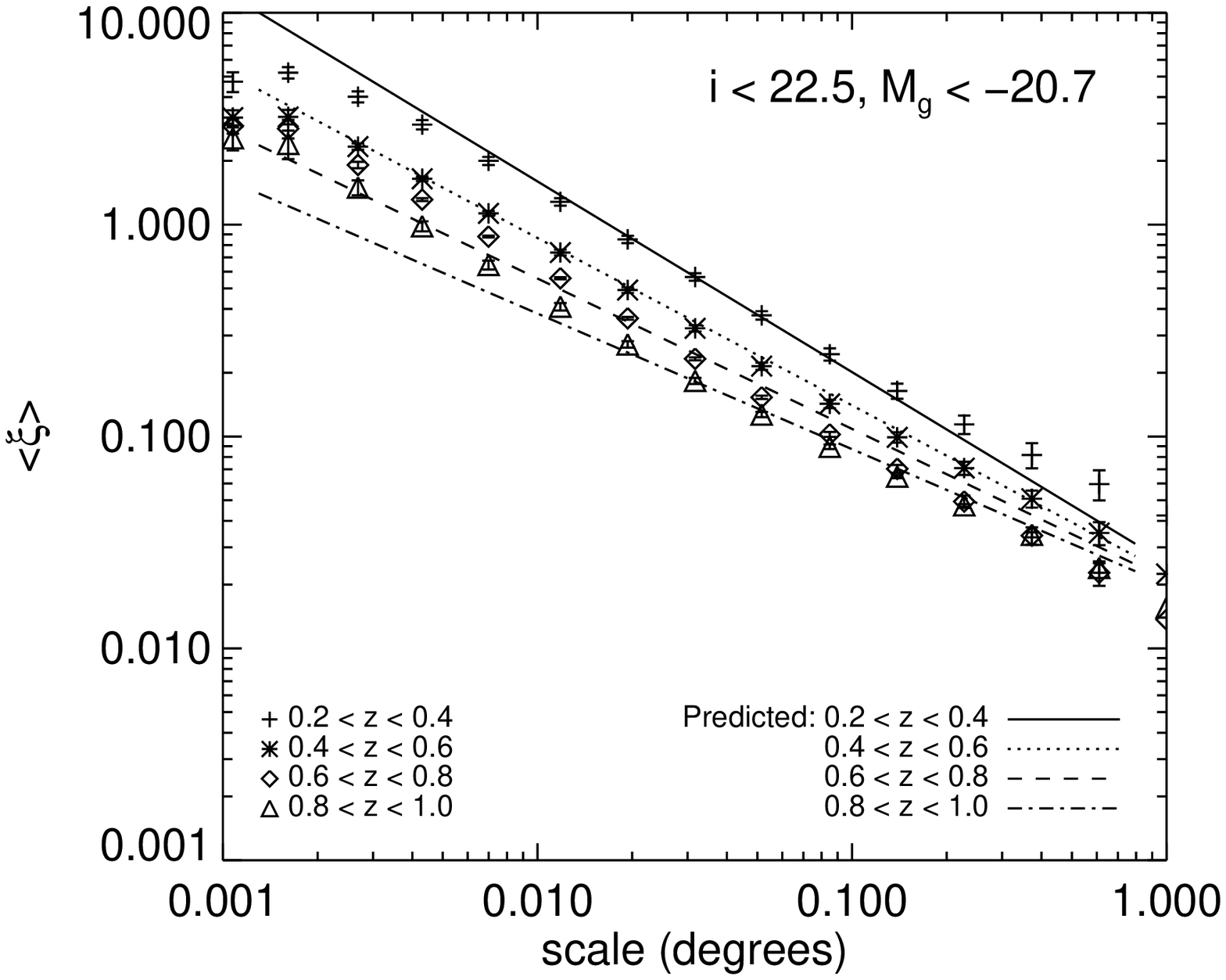}
      \end{tabular}
      \end{center}
      \caption{The symbols on the left panel represent the measurements of
        $w(\theta)$ in the four redshift bins. The lines are the fits obtained
        using a $\chi^{2}$ assuming a power law for the two point correlation
        function. The symbols on the right panel show the averaged two point correlation
        measured using counts in cells. The lines are the predicted
        values using formula (31) with the parameters of our $w(\theta)$ fit.}
\label{figure:predict}
\end{figure*}

Considering the average two point correlation (defined by equation
\ref{eq:xiavdef}) on Figure~\ref{figure:predict}, we see, as
expected, the same behaviour as $w(\theta)$ i.e that the clustering
increases as the redshift decreases. We can also use this measurement
to verify that the Landy \& Szalay estimator measurements are
consistent with BMW-PN measurements. To do this, we fit $w(\theta)$ by
a simple power law $w=A\theta^{1-\gamma}$ using a $\chi^{2}$ to
estimate the amplitude $A$ and the power $\gamma$. Our results are
plotted on the left panel of Figure \ref{figure:predict}.

Using the formula \ref{eq:xiavdef} with square cells, we derive an
approximate analytical expression for the average two point
correlation function:
\begin{equation}
\wbar \sim \frac{4 \times 2^{(3-\gamma)/2}}
{(3-\gamma)(4-\gamma)} \, A \theta^{1-\gamma}.
\end{equation}

Comparing with a numerical integration, we found that this formula
works for $\gamma$ in the range $[1.6,2.0]$ (our values of $\gamma$
are $1.9$, $1.8$, $1.7$ and $1.65$ for the redshift bins $0.2<z<0.4$,
$0.4<z<0.6$ , $0.6<z<0.8$ and $0.8<z<1.0$ respectively) with a
precision that increases with $\gamma$. The differences in $\wbar$
between the two methods ranges from $8.5\%$ for $\gamma=1.6$ to
less than $1\%$ for $\gamma=2$. The results are presented on the right
panel of Figure \ref{figure:predict}. We can see that this prediction
matches very well the measurements.

%\begin{figure}
%\includegraphics[width=7cm]{combinedxiav.eps}
%\caption{Combined $\overline{\xi}$ as a function of scale in the four
%  redshift bins. The error bars are field-to-field error bars. We see
%  that as expected the clustering amplitude is higher at low redshifts.}
%\label{figure:xiav}
%\end{figure} 

\subsection{Higher-order moments}

\subsubsection{Comparison with SDSS DR7}
\label{sec:comparison-with-sdss}

We now consider our measurements of the hierarchical moments. As first
consistency check, we compare our $S_n$s measured in the CFHTLS to
measurements made from the seventh data release of the Sloan Digital
Sky Survey (SDSS-DR7) using an independently-written counts-in-cells
code \cite{RBM07}. This code was used with the default configuration
parameters proposed by the authors.

\begin{figure}
\includegraphics[width=8cm]{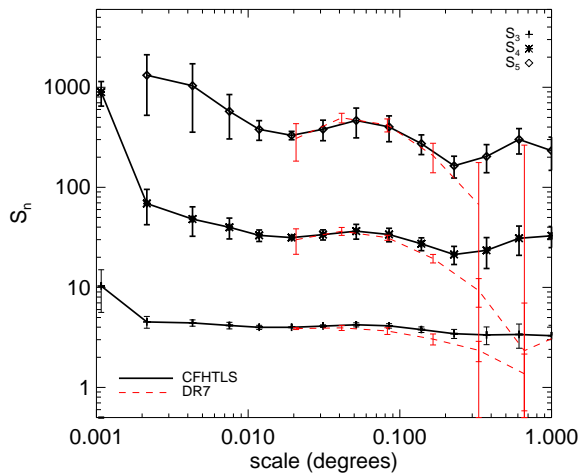}
\caption{Hierarchical moments $S_3$, $S_4$ and $S_5$ measured in the
  DR7 release of the SDSS for galaxies selected with $0<z<0.4$, $18<r<21$
and $M_{r}<-20.7$ (dashed lines) compared to measurements in the
CFHTLS (solid lines).}
\label{figure:SDSS}
\end{figure} 

Since the SDSS and CFHTLS are by their nature very different surveys,
we can only make this comparison for the lowest redshift bins in the
CFHTLS.  We therefore chose galaxies with $0<z<0.4$, $18<r<21$ and
$M_{r}<-20.7$ \citep[comparable to][]{RBM07} and applied the same
selection to galaxies in the SDSS. Our results are plotted in Figure
\ref{figure:SDSS}, where we show the hierarchical moments obtained
from the four fields of the CFHTLS (solid lines) compared to
measurements in the SDSS (dashed lines). For the CFHTLS, error bars
are computed according to the method described in
\S~\ref{sec:comb-heir-momem}; for the SDSS, error bars were computed
using a jackknife method \citep[e.g.,][]{Scranton02}. For the detailed
method see \cite{RB06}.

As a consequence of the extremely fine pixelisation grid we employ, we
are able to measure our $S_n$ to much smaller angular scales than in
the SDSS (such a measurement would be computationally demanding for a
large-area survey like the SDSS).  In the common angular range, the
agreement between the CFHTLS and the SDSS measurements is excellent
except at the largest scales. In this latter regime, however, our
measurements are expected to be less precise, due to finite size
effects when approaching the angular size of the fields.

\subsubsection{Field-to-field variations}

\begin{figure*}
  \begin{center}
  \hbox{
      \includegraphics[width=8cm]{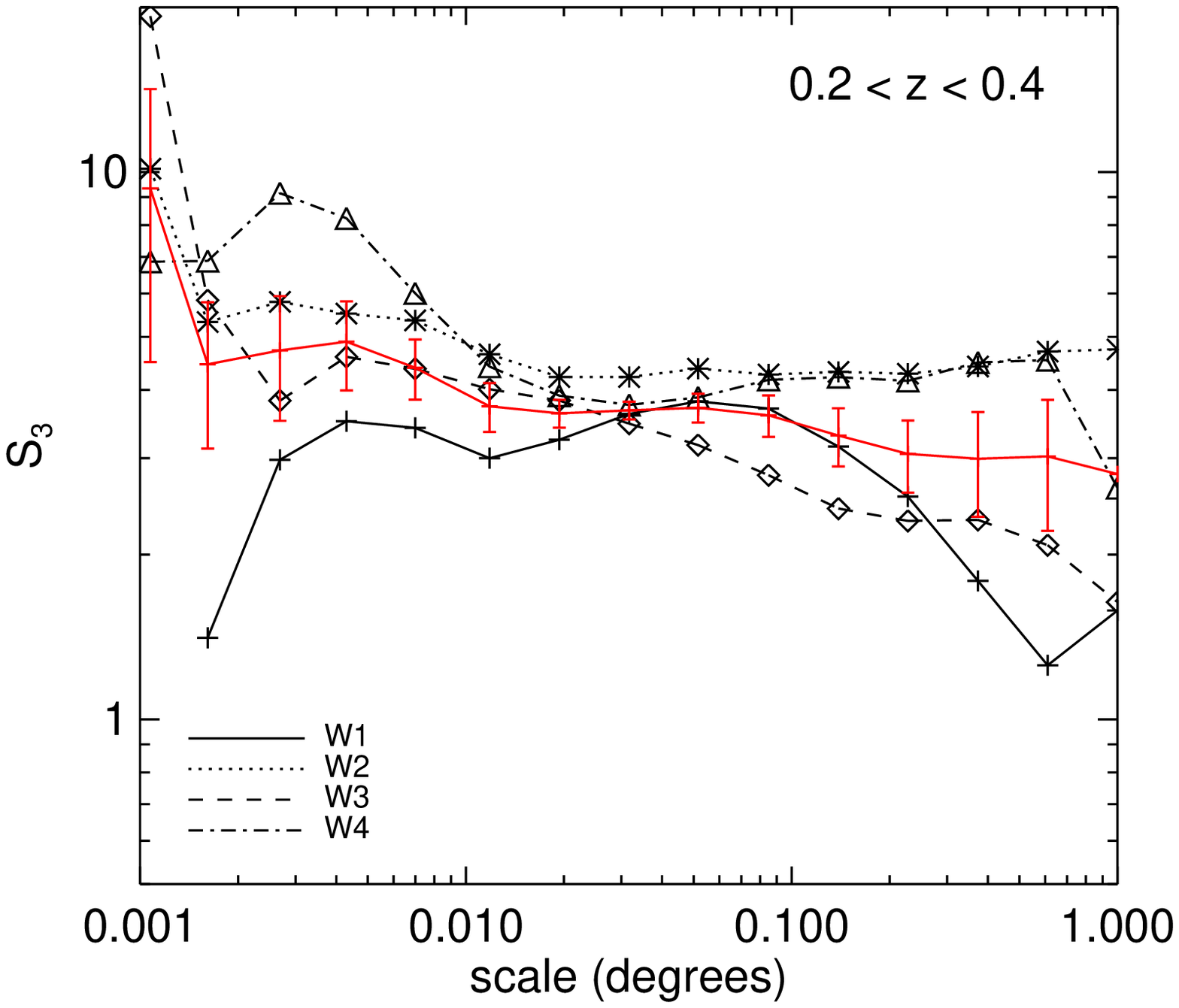}
      \includegraphics[width=8cm]{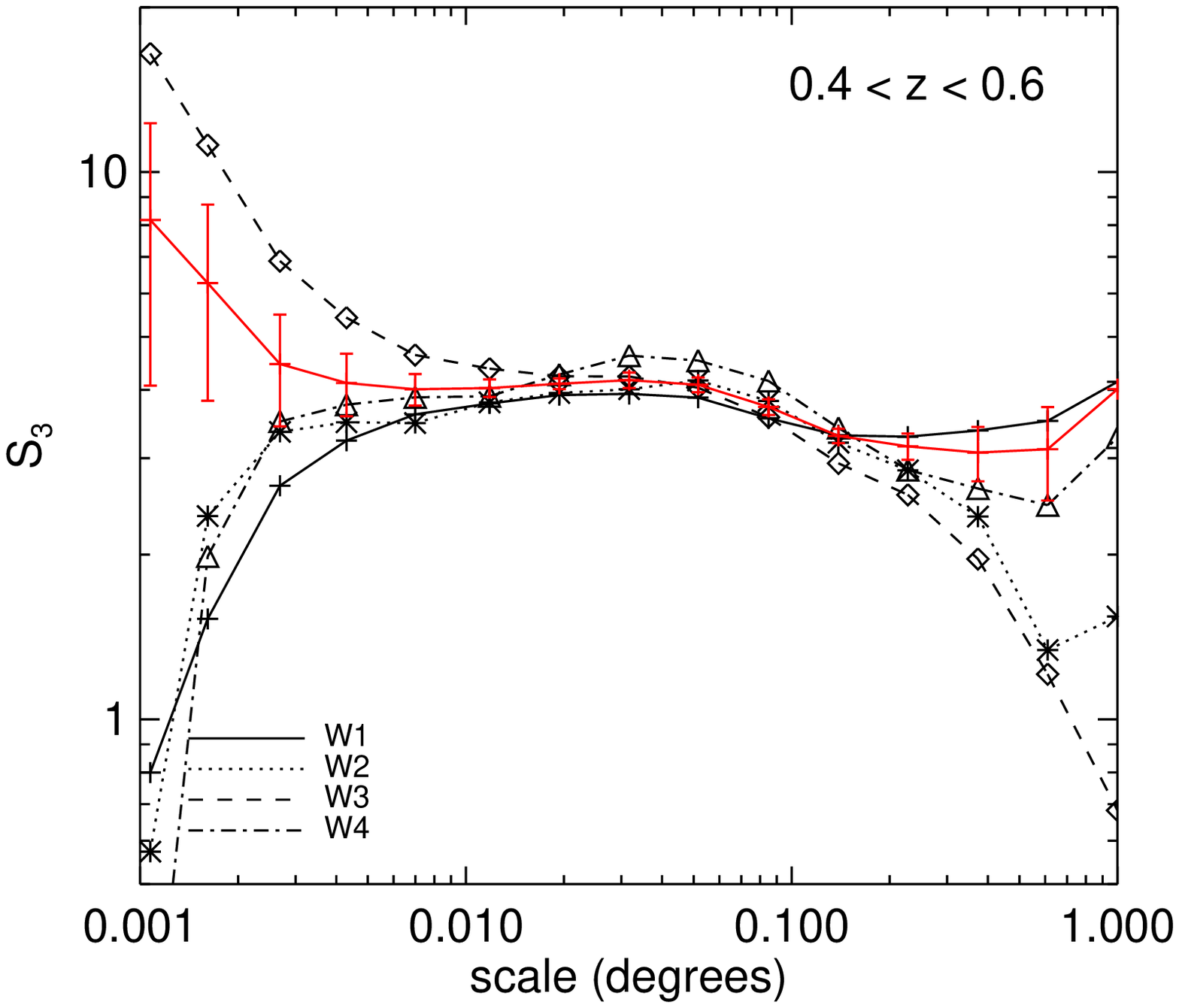}}
\hbox{
      \includegraphics[width=8cm]{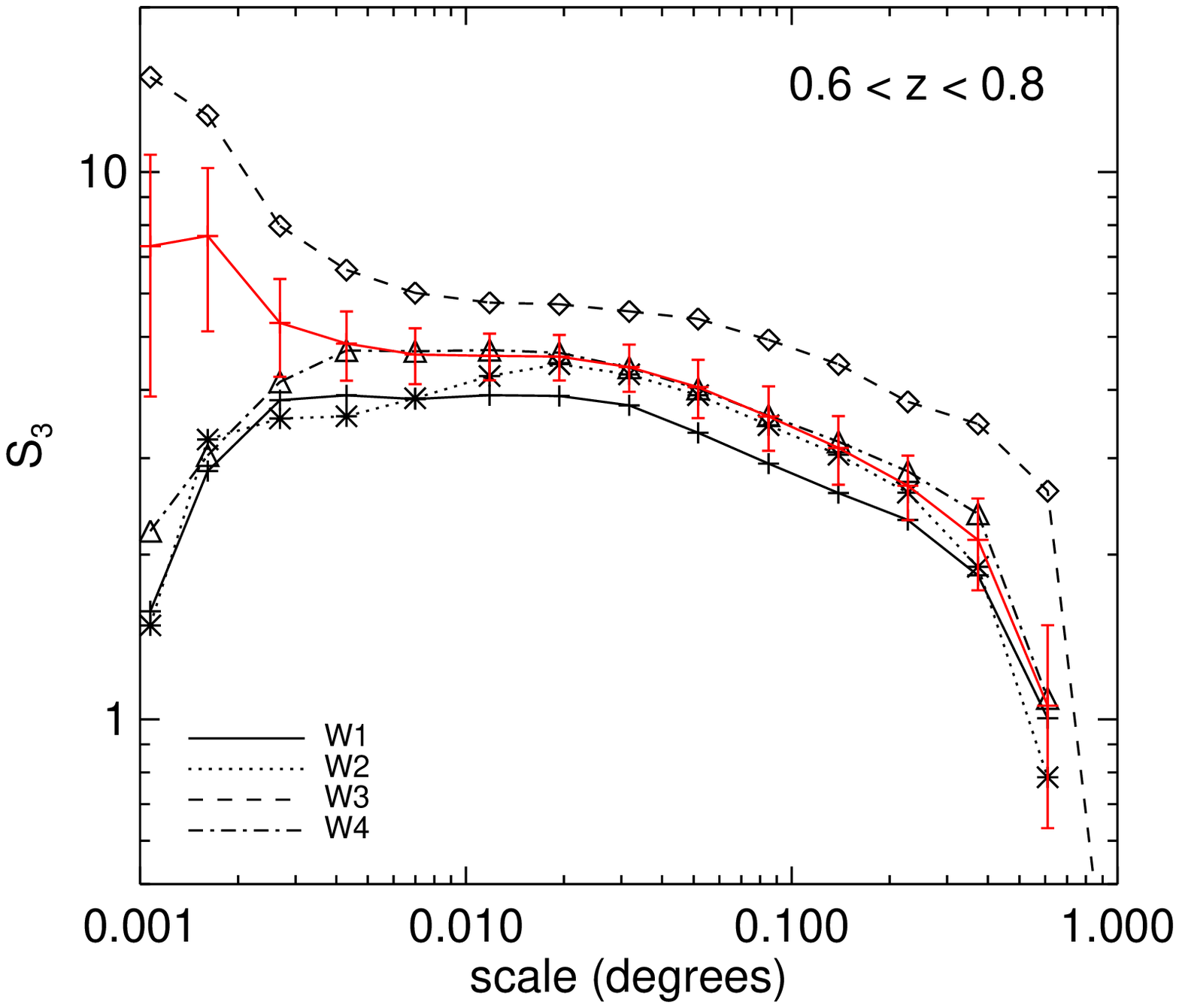}
      \includegraphics[width=8cm]{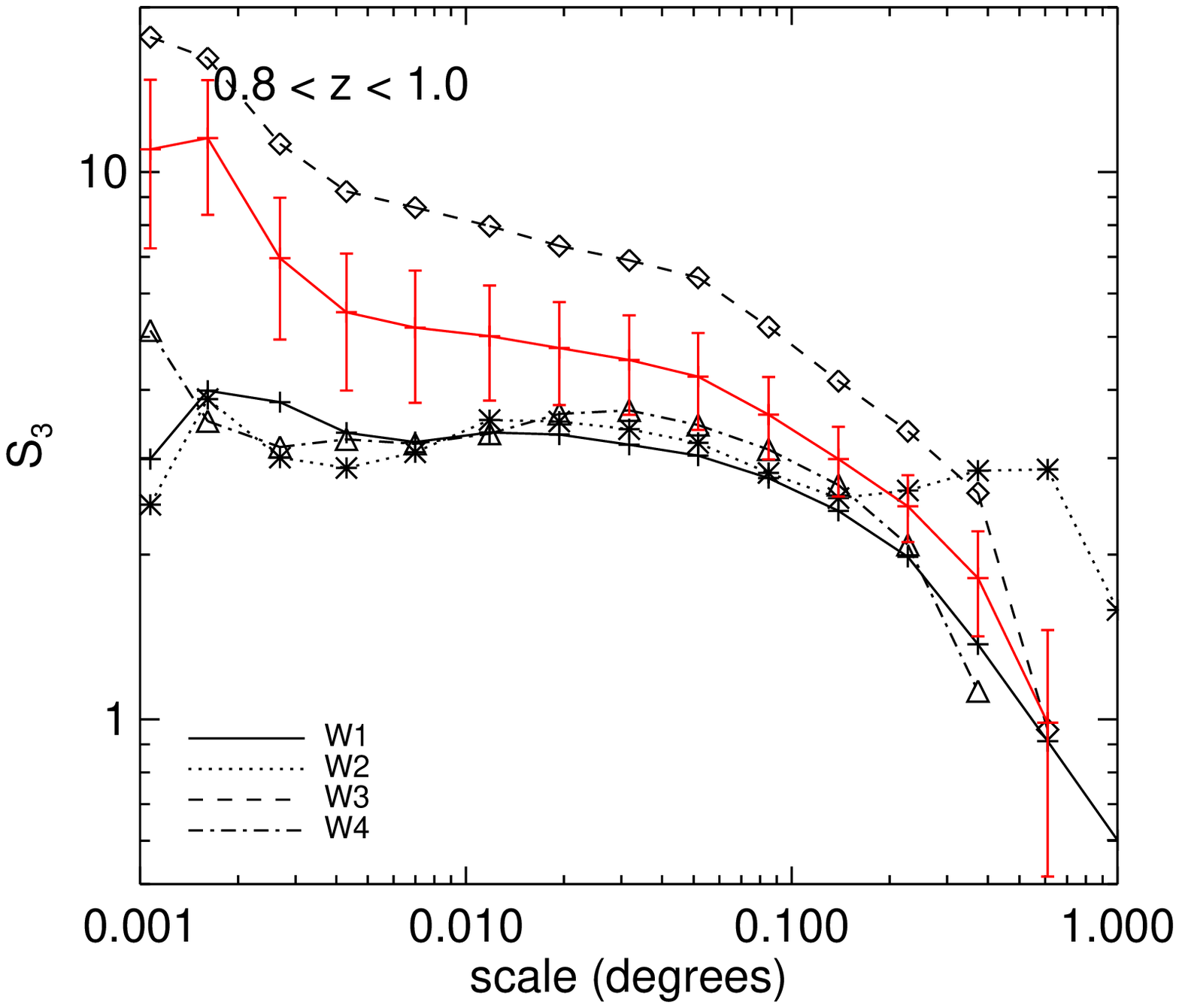}}
     
      \end{center}
      \caption{The four panels represent measurements of $S_{3}$ in
        the four fields of the CFHTLS from the lowest redshift bin
        $0.2<z<0.4$ to the highest one $0.8<z<1.0$.  The red line
        represents the combined $S_{3}$ computed using the method
        described in Section~\ref{sec:comb-heir-momem}. We can see
        that at low redshift all the fields behave identically while
        at higher $z$, $S_{3}$ in W3 is systematically higher over the
        whole scale range.}
\label{figure:S3fields}
\end{figure*}

\begin{figure*}
  \begin{center}
 
      \includegraphics[width=8cm]{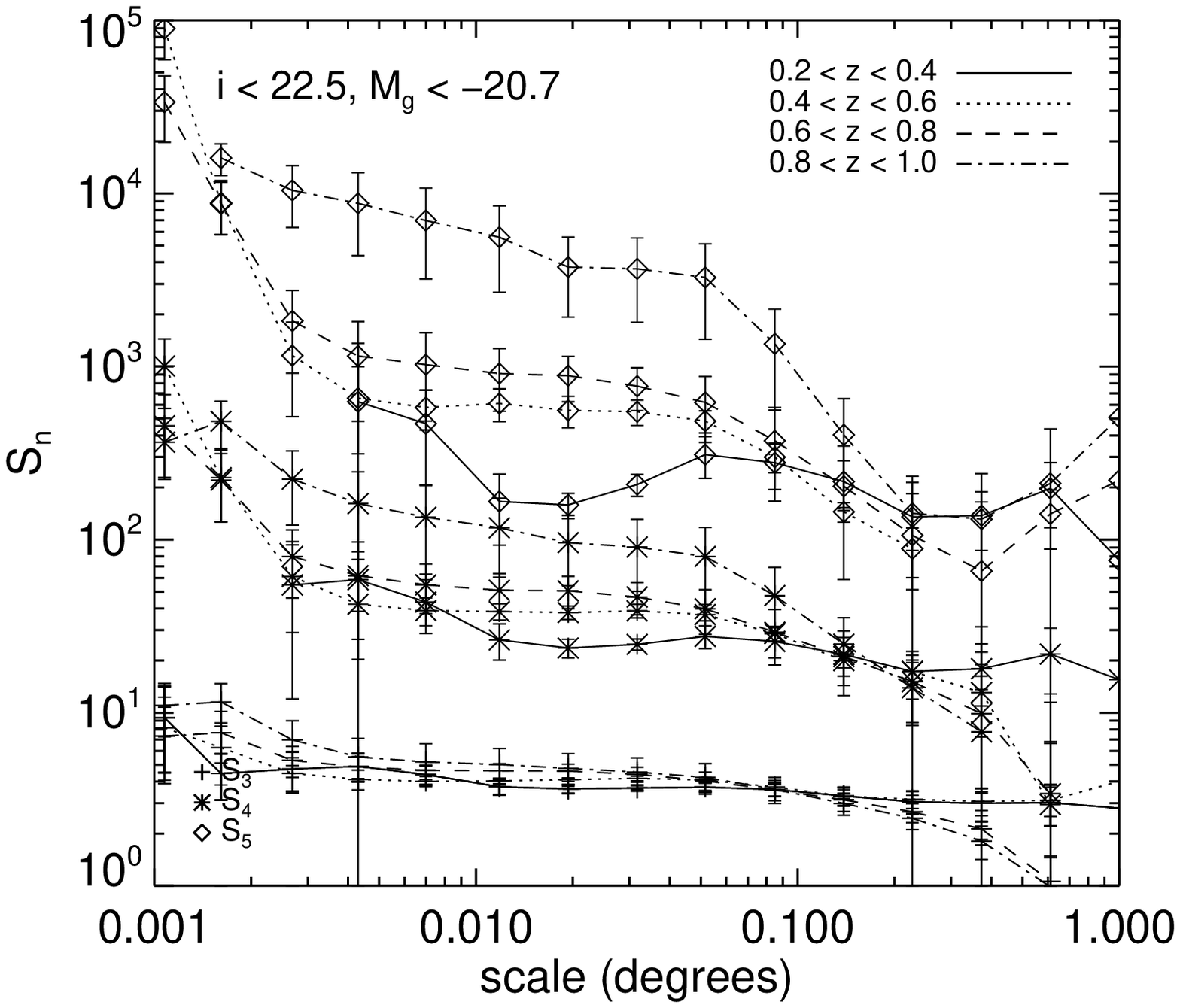}
      \includegraphics[width=8cm]{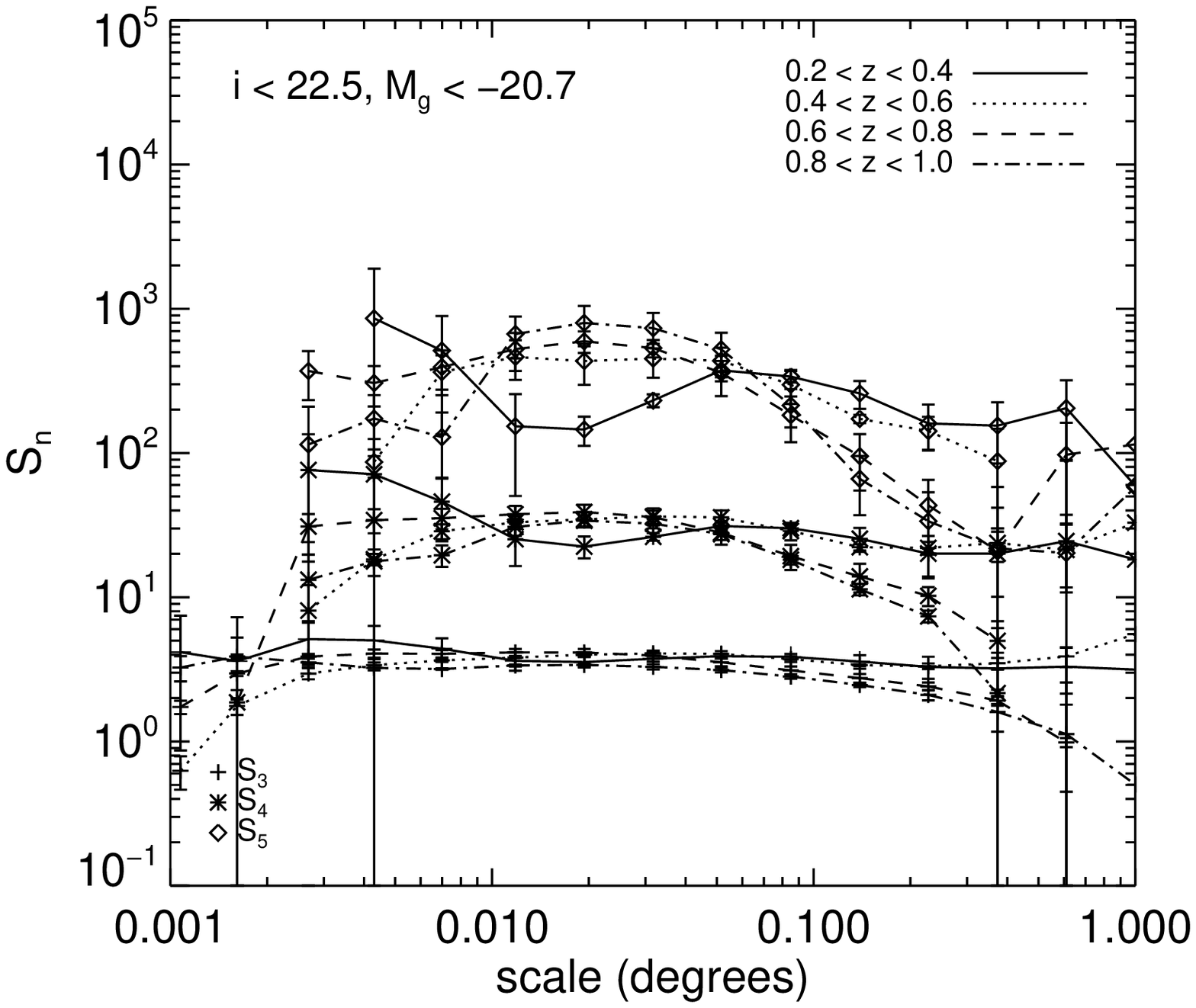}

      \end{center}
      \caption{Combined $S_{n}$'s as functions of
        angular scale in the four redshift bins when using W1+W2+W3+W4
        together (left panel) and when W3 is excluded  (right
        panel). Points which are not plotted correspond to negative values.}
\label{figure:Sns}
\end{figure*}

Figure~\ref{figure:S3fields} shows the third-order cumulant $S_{3}$ as
a function of scale in different redshift bins for the four different
fields. At low redshifts, $z \la 0.6$, the dispersion between the four
fields is low; however, above $z \simeq 0.7$ W3 begins to be
systematically higher than the other fields at all scales. At the
highest redshifts, the difference between W3 and the other fields is
significant. The same behaviour is observed for the other cumulants. To
illustrate furthermore this point Figure~\ref{figure:Sns} compares the
hierarchical moments as a function of scale for the various redshift
bins for the cases when all the fields are combined together and when
W3 is excluded.  The presence of W3 is in fact crucial because when it
is included there clearly seems to be a redshift evolution of the
hierarchical moments as indicated by the left panel of
Figure~\ref{figure:S3fields}, while removing W3 makes the redshift
dependence insignificant as shown on the right panel. We now
investigate this effect in more detail. 

\begin{figure*}
  \begin{center}
    \includegraphics[width=13cm]{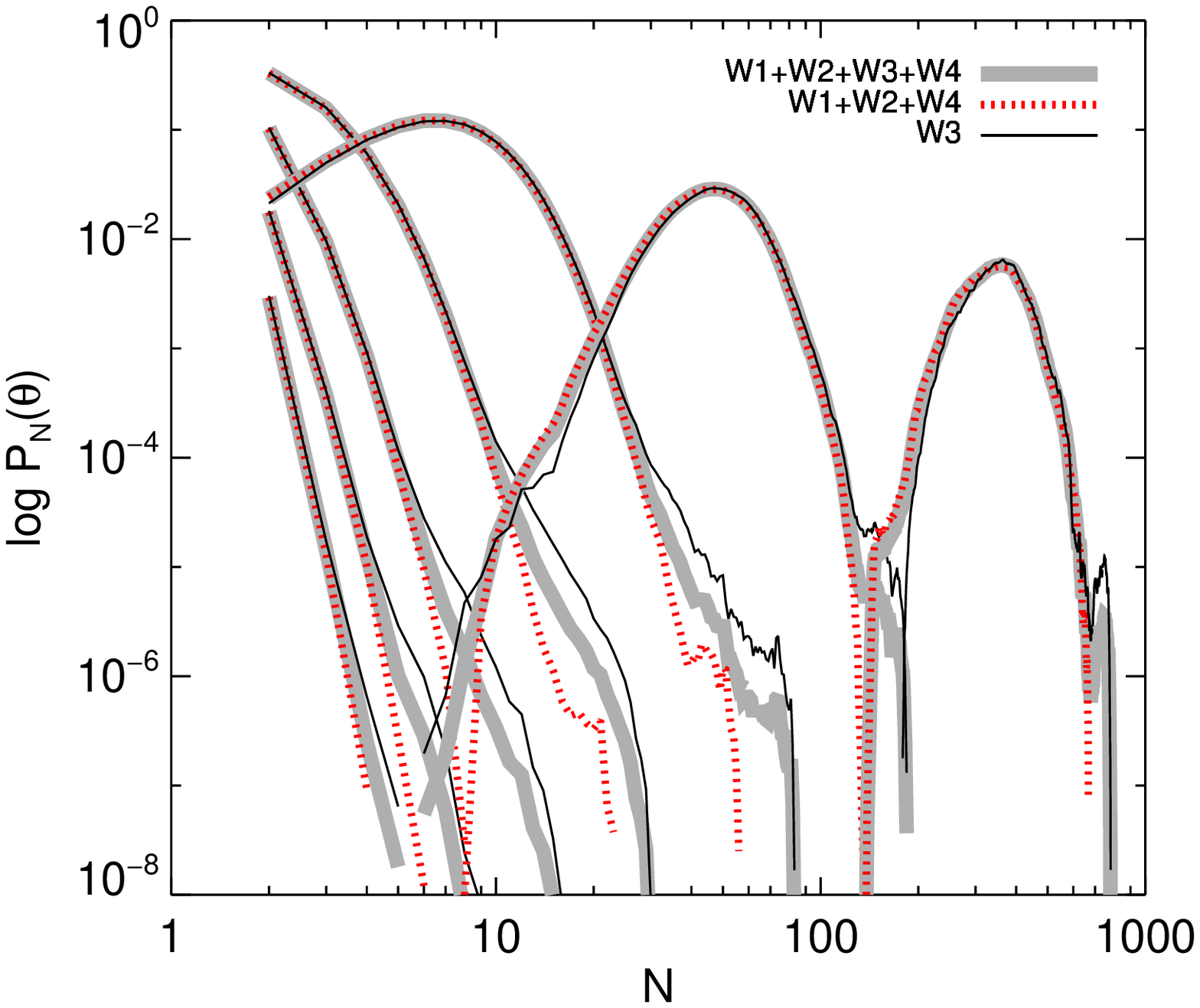}
  
  \end{center}
  \caption{Counts in cells distribution function $P_{N}(\theta)$ as a
    function of $N$ in the last redshift bin $0.8<z<1.0$. Three cases
    are considered: the combination W1+W2+W3+W4 using the weighting
    scheme as in equation (\ref{eq:combined}) (thick grey curves), the
    combination W1+W2+W4 (red dotted curves) and W3 alone (solid black
    curves). Each curve corresponds to a different scale; for clarity
    we keep only one scale out of two. }
\label{figure:PofN} 
\end{figure*}

The explanation of the behavior of the cumulants at the highest
redshift bin can be understood from an examination of Figure
\ref{figure:PofN}, which shows the counts in cells distribution
function for the combined fields, W1+W2+W3+W4, for W1+W2+W4, and for
W3 alone. Clearly, the presence of W3 dominates and perturbs
considerably the high end tail of $P_{N}(\theta)$. A careful
inspection of the W3 field reveals that this is due to the
presence of one or several rich clusters in W3 compared to the three
other fields \citep[see e.g.][]{Colombi94}. Due to the finiteness of
the sample there is a value $N_\max(\theta)$, above which
$P_{N}(\theta)$ becomes zero. This function $N_\max(\theta)$ obviously
relates to the richest clusters of the sample: for instance a cell of
size $\theta$ located in the densest part of one of the richest
cluster of the sample should contain $N_\max(\theta)$
galaxies. Similarly, $P_{N}(\theta)$ for $N$ close to $N_\max$ is
entirely determined for a given smoothing scale $\theta$ by the
density profile of a rich cluster of galaxies \citep[cf. Figure 5
of][]{Colombi94}.

\begin{figure*}
  \begin{center}
    \hbox{
      \includegraphics[width=8cm]{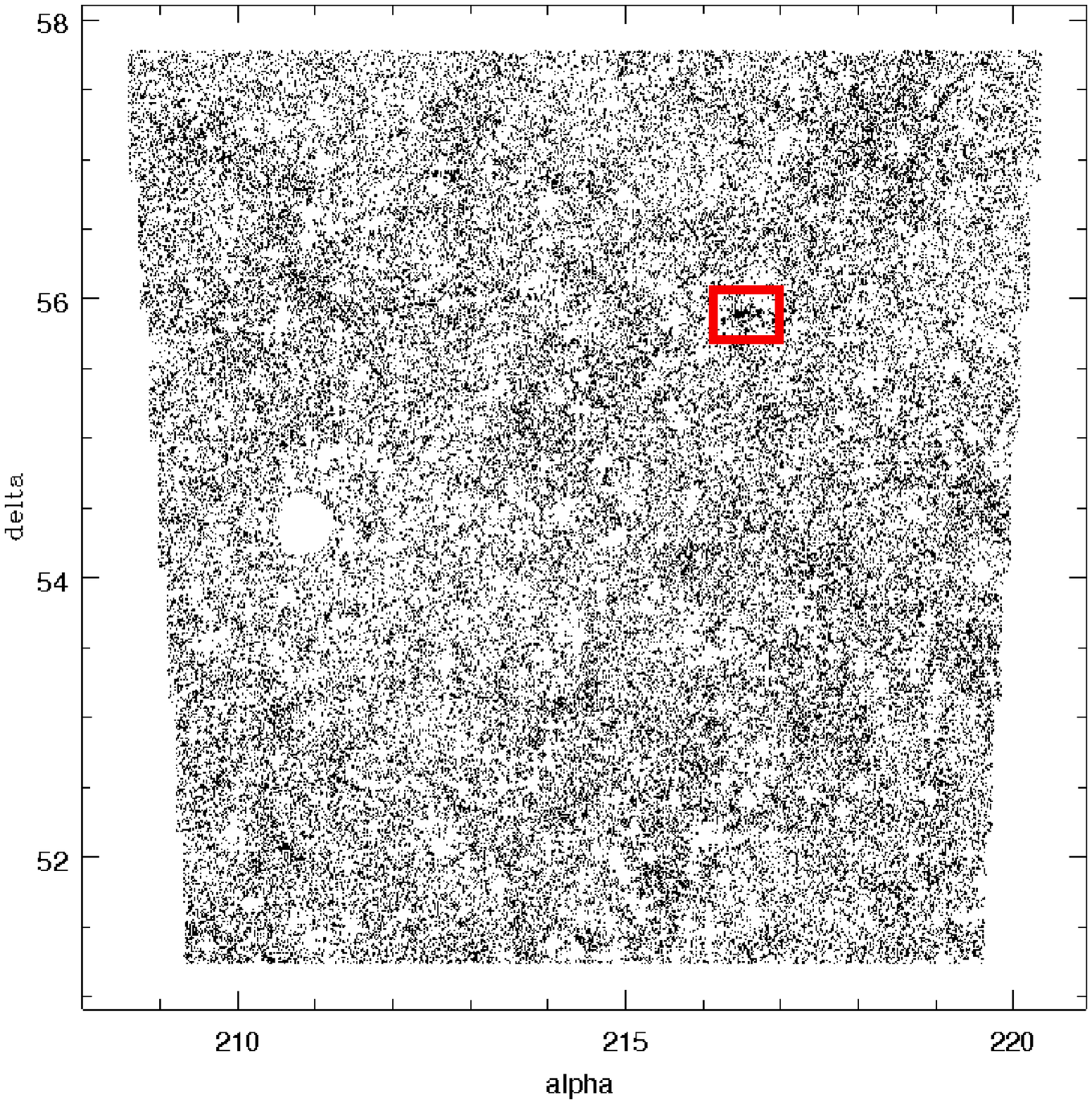}
      \includegraphics[width=9cm]{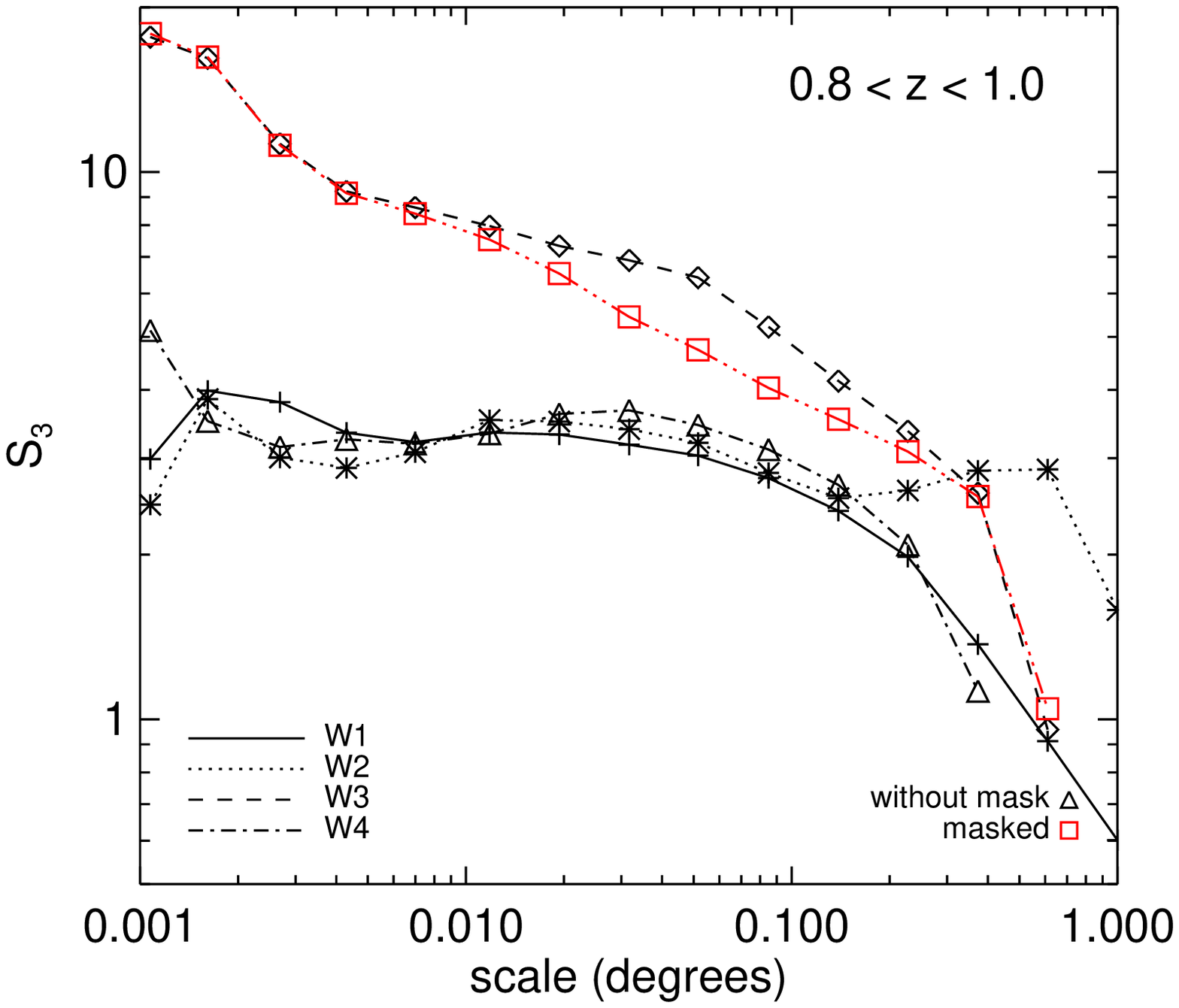}}
    \hbox{
      \includegraphics[width=8cm]{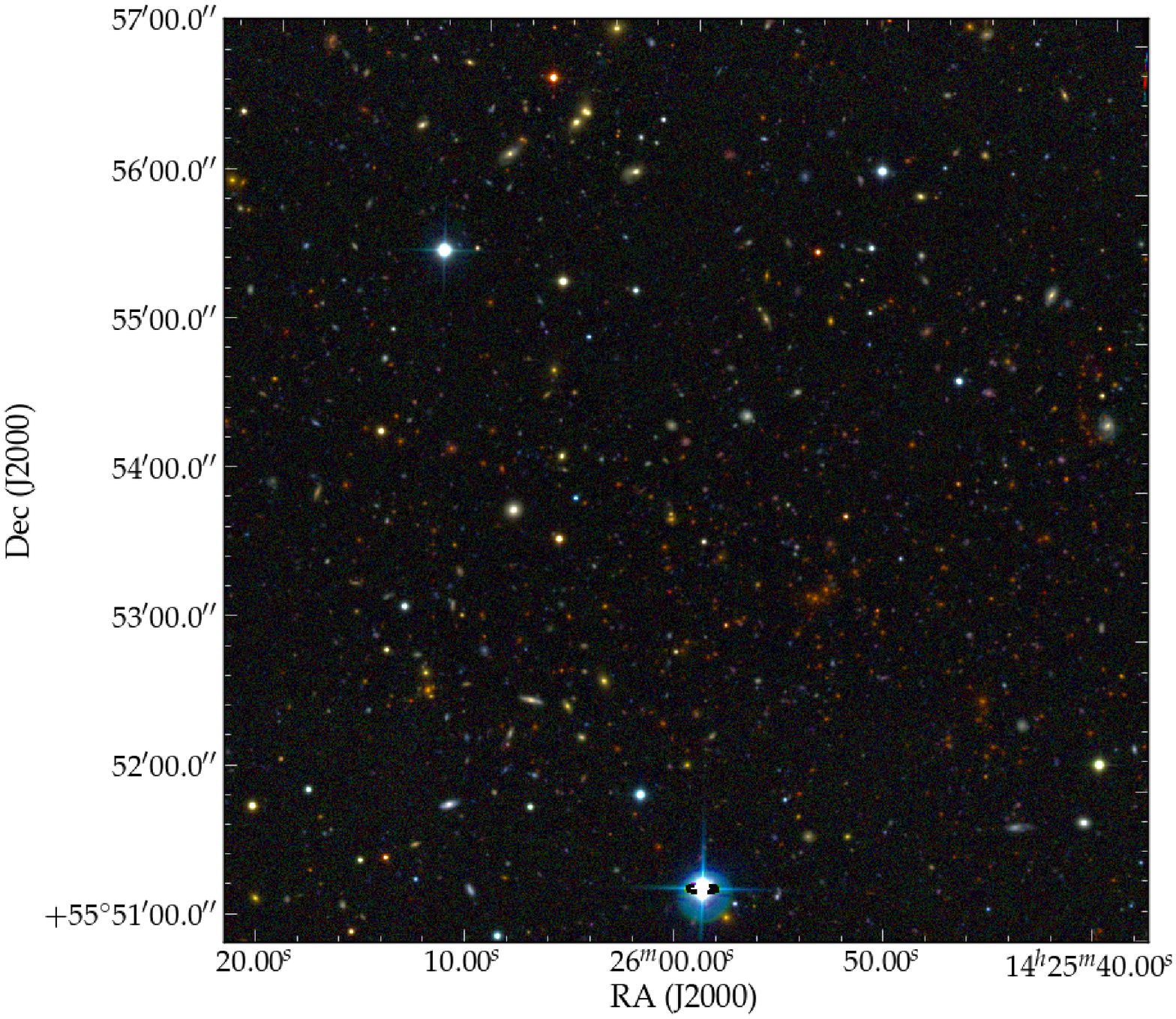}
      \includegraphics[width=9.8cm]{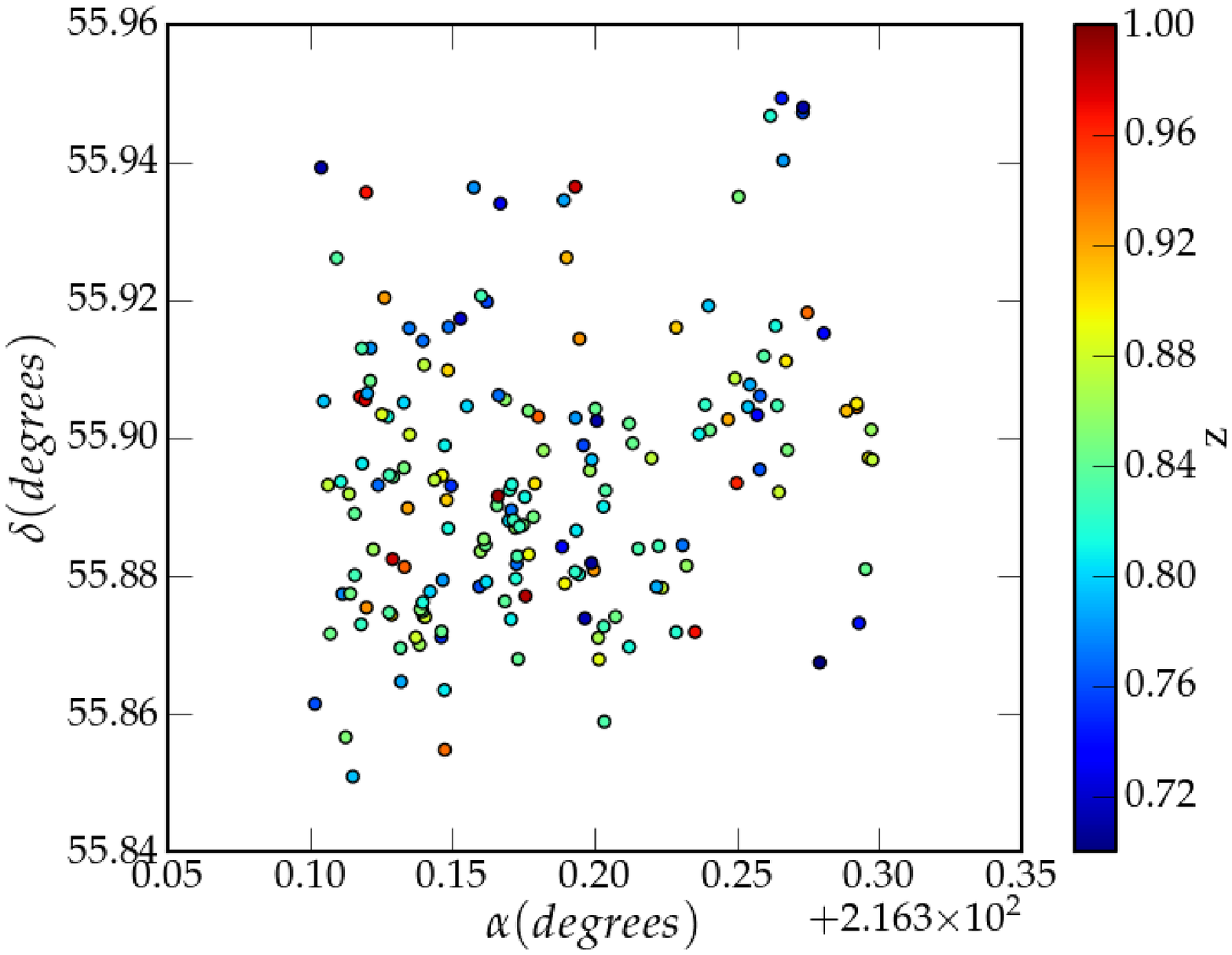}}
  
  \end{center}
  \caption{A rich cluster in W3. The upper left panel represents the
    distribution on the sky of galaxies in W3 for $0.8<z<1$. The red
    square encloses the most prominent concentration from a visual
    inspection. The upper right panel illustrates the effect of
    masking this concentration on $S_{3}$ for W3 in the highest
    redshift bin.  The lower left panel shows a coloured image of the
    sky in the region delimited by the red square, which reveals a
    cluster of red galaxies. On the lower right panel, the galaxies of
    the same region are represented in the same field with symbol
    colour representing redshift. We use a wider redshift bin
    ($0.7<z<1.0$) to be sure not to cut galaxies that belong to the
    cluster due to photometric redshift errors.}
  \label{figure:posW3}
\end{figure*}

To illustrate furthermore this point, Figure \ref{figure:posW3} shows
the largest cluster in W3 at $0.8<z<1.0$ \footnote{this cluster has
  coordinates $(216.5,55.9)$ in radec J2000.} and the effect on
$S_{3}$ of subtracting it from the field. As expected, removing the
cluster greatly reduces the skewness over a large angular
range. Indeed, moments of order $k$ of the count probability are
increasingly sensitive to the high-$N$ tail of $P_{N}(\theta)$ with
increasing $k$. Removing the largest cluster reduces the high $N$ tail
of $P_{N}(\theta)$ and hence reduces the skewness. This happens here
only on a finite range of angular scales because, at smaller (larger)
scales, one or several clusters with different concentration
parameters from those of the one we selected dominate the high $N$
tail of the count probability.  In fact, detailed examination of the
four fields in the redshift bin $0.8<z<1$ shows that there are four
rich clusters in W3 and one in W1, which affect the cumulants and explains the
discrepancy between W3 and the other fields.

\subsubsection{Significance of field-to-field variations}
\label{sec:comb-meas}

To test if the discrepancy between W3 and the three other fields is
statistically significant we decided to compute the theoretical
expectation of the errors on $\bar{\xi}$, $S_{3}$ and $S_{4}$ in order
to compare them to the errors estimated as described in
\S~\ref{sec:comb-heir-momem} and as shown in
Table~\ref{table:Sns}. This is made possible by the use of the package
{\tt FORCE} \citep[FORtran for Cosmic Errors,][]{Colombi01} which
exploits the analytical calculations of \cite{Szapudi96} and
\cite{Szapudi99} to provide quantitative estimates of these
statistical errors due to the finiteness of the sample, given a number
of input parameters that specify a prior model.

All the parameters needed for such a calculation were estimated from
the survey self-consistently. These are the area $S$ of the survey,
the cell area $s$, the average galaxy number count $\bar{N}$, the
averaged two-point correlation function over a cell $\bar{\xi}$, the
values of $S_{n}$'s for $3 \leq n \leq 8$ and the averaged two-point
correlation function over the survey, \beq \xibar_{S}=\frac{1}{S^{2}}
\int_{S} \,\d\theta_{1}\,\d\theta_{2} \, w(\theta_{1},\theta_{2}).
\label{eq:intS}
\eeq Except for $\xibar_{S}$, all the parameters above can be
extracted directly from the data, including $S_{n}$, which measured up
to the eighth order.

The calculation of $\xibar_{S}$ is however more complex.  To perform
it we use linear perturbation theory to derive a three-dimensional
correlation function for a ${\Lambda}$CDM model. Then, we compute the
angular two point correlation function in the various redshift bins
using equation (\ref{w2}), from which we can derive numerically the
integral (\ref{eq:intS}).\footnote{$w(\theta)$ was computed using the
  halo-model module of {\tt COSMOPMC (http://cosmopmc.info)}, while
  $\xibar_{S}$ was estimated by Monte-Carlo simulation using $10^{6}$
  pairs to perform the integral.} The result is multiplied with the
square of the bias between the galaxy and the dark matter
distributions. To estimate this linear bias, we fit the measured
two-point correlation function using the ``halo-model'' \citep{SSHJ01,
  Maetal00, Peacock00, Coorayetal02}, exactly as in
\cite{Couponetal12}.\footnote{using again {\tt COSMOPMC}.}  The values
are given in Table~\ref{table:biasvalue}.

\begin{table}
\centering
\begin{tabular}{cccccccc}
\hline
Redshift bin & $b$ \\
\hline
$0.2<z<0.4$ & 1.17 \\
$0.4<z<0.6$ & 1.21 \\
$0.6<z<0.8$ & 1.27 \\
$0.8<z<1.0$ & 1.37 \\
\hline
\end{tabular}
\label{table:biasvalue}
\caption{Bias in the different redshift bins.}
\end{table}

\begin{figure*}
  \begin{center}
    \begin{tabular}{c@{}c@{}}
      \includegraphics[width=5.5cm]{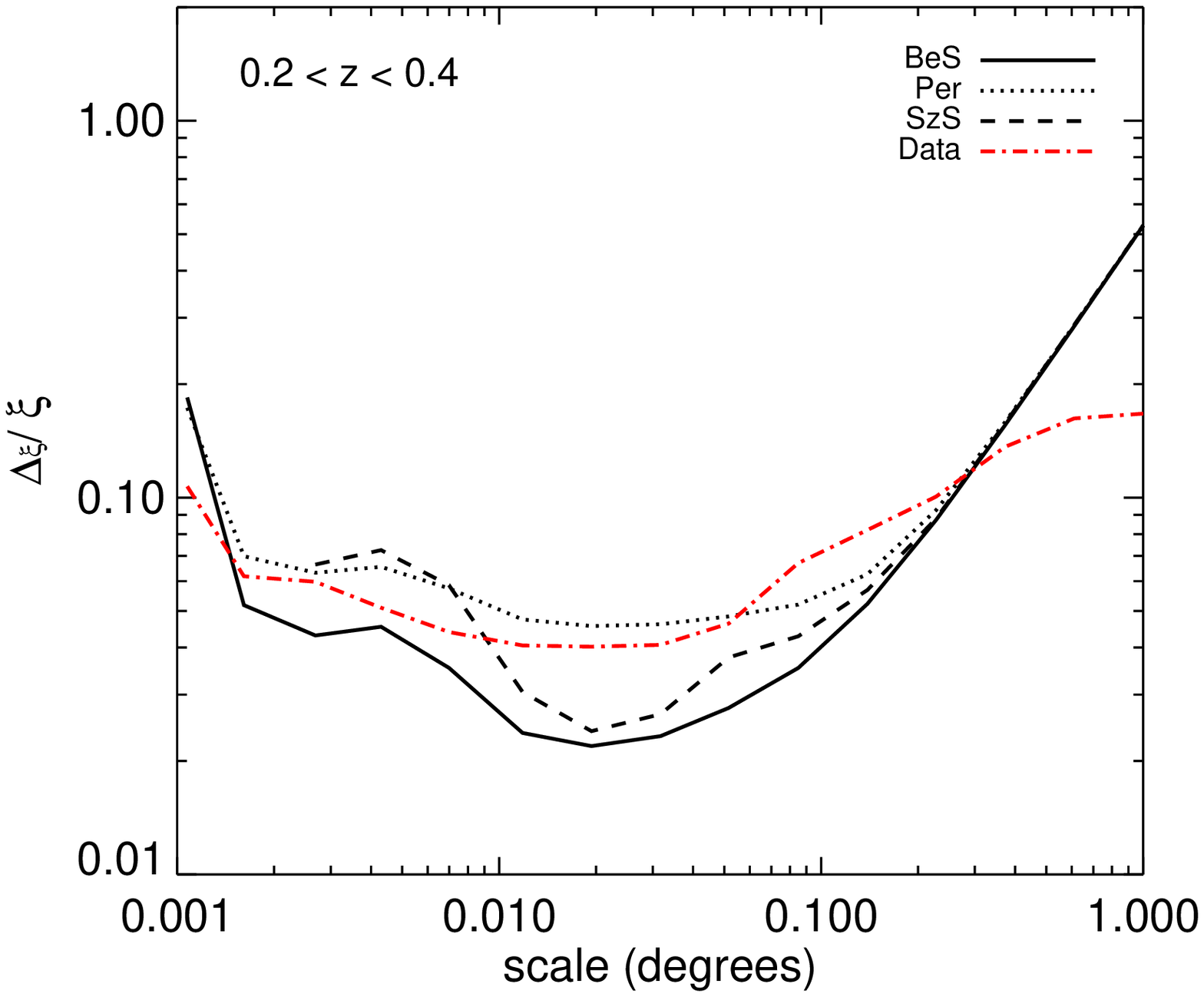}
      \includegraphics[width=5.5cm]{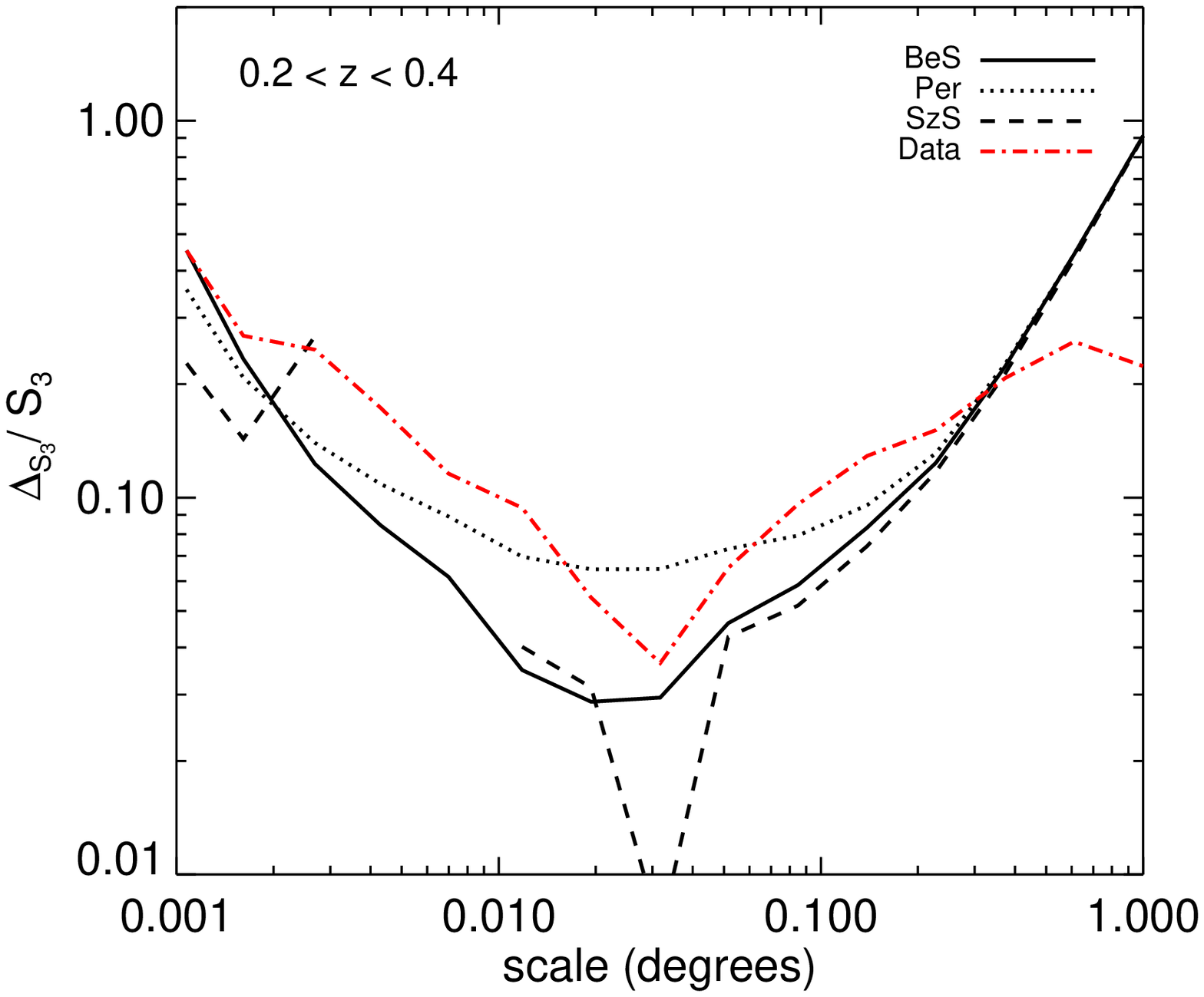}
      \includegraphics[width=5.5cm]{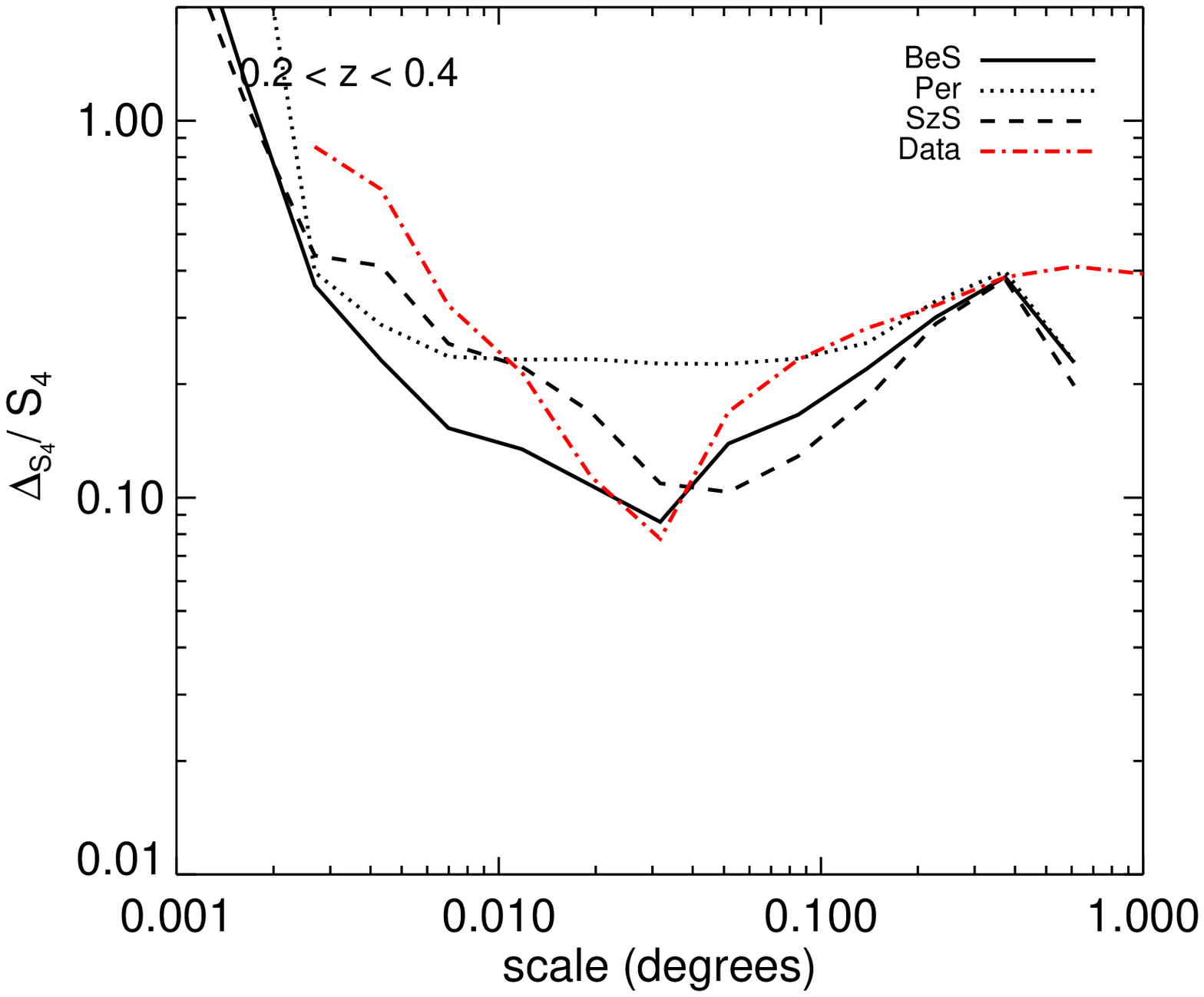}
      \end{tabular}
      \begin{tabular}{c@{}c@{}}
      \includegraphics[width=5.5cm]{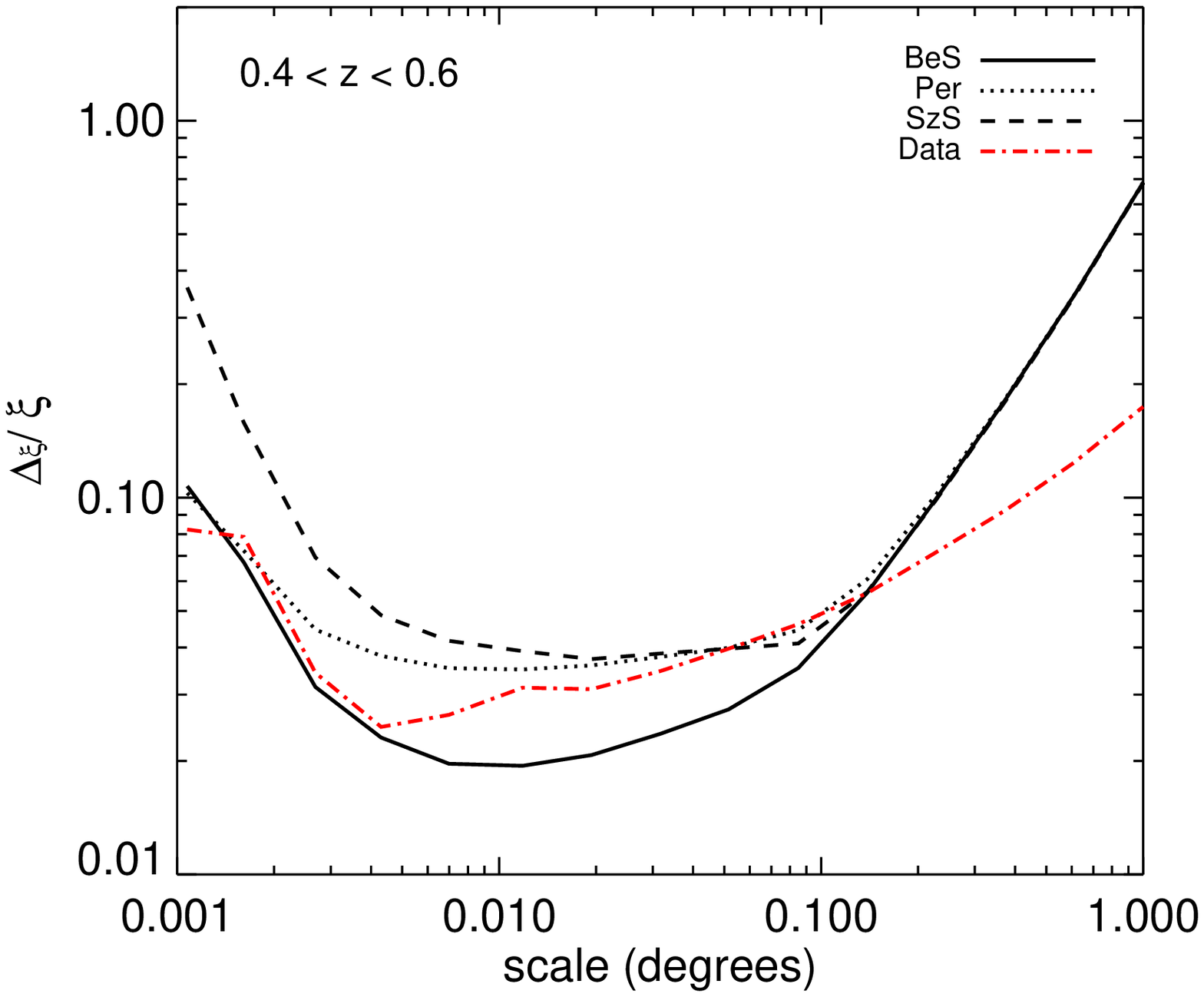}
      \includegraphics[width=5.5cm]{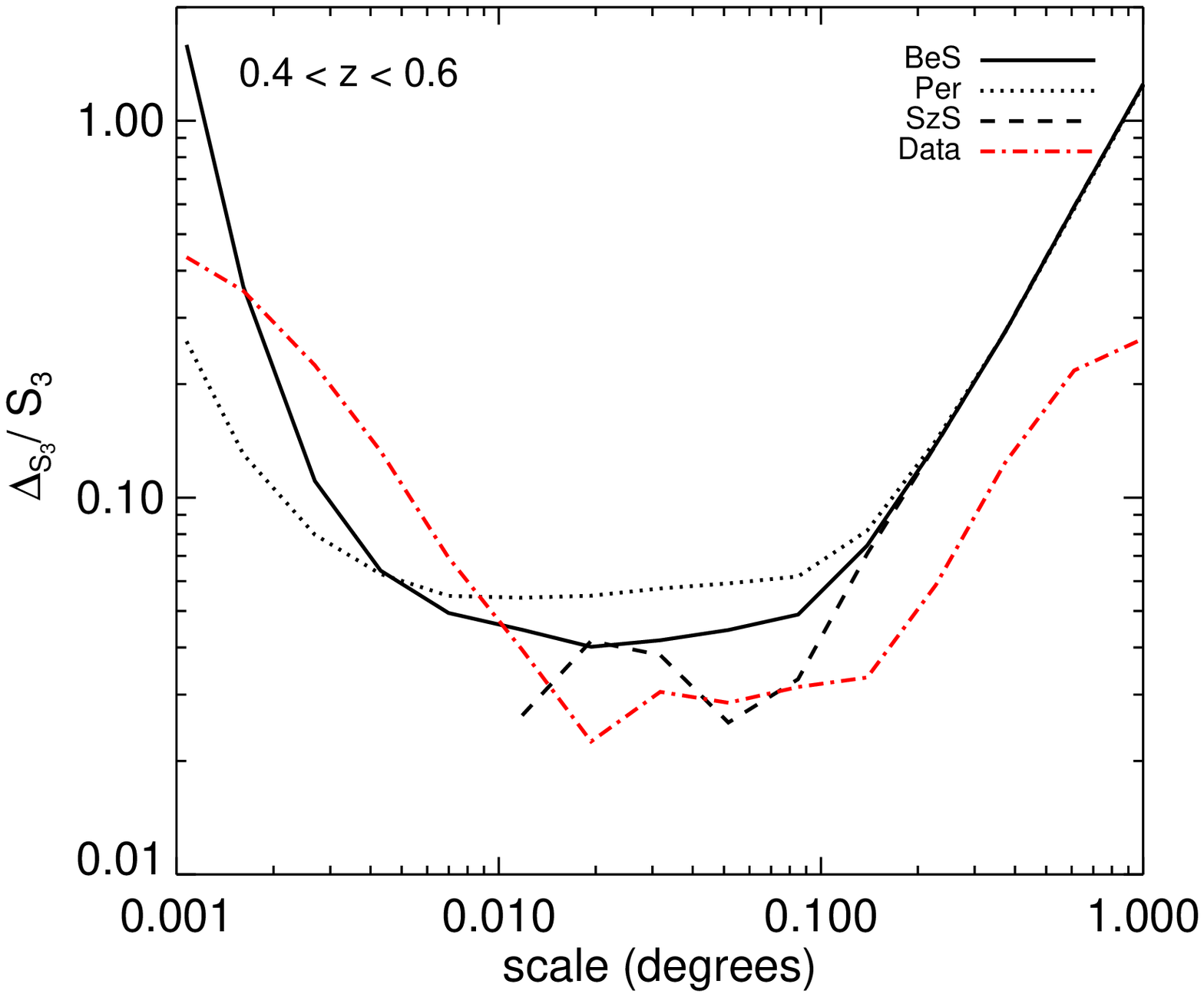}
      \includegraphics[width=5.5cm]{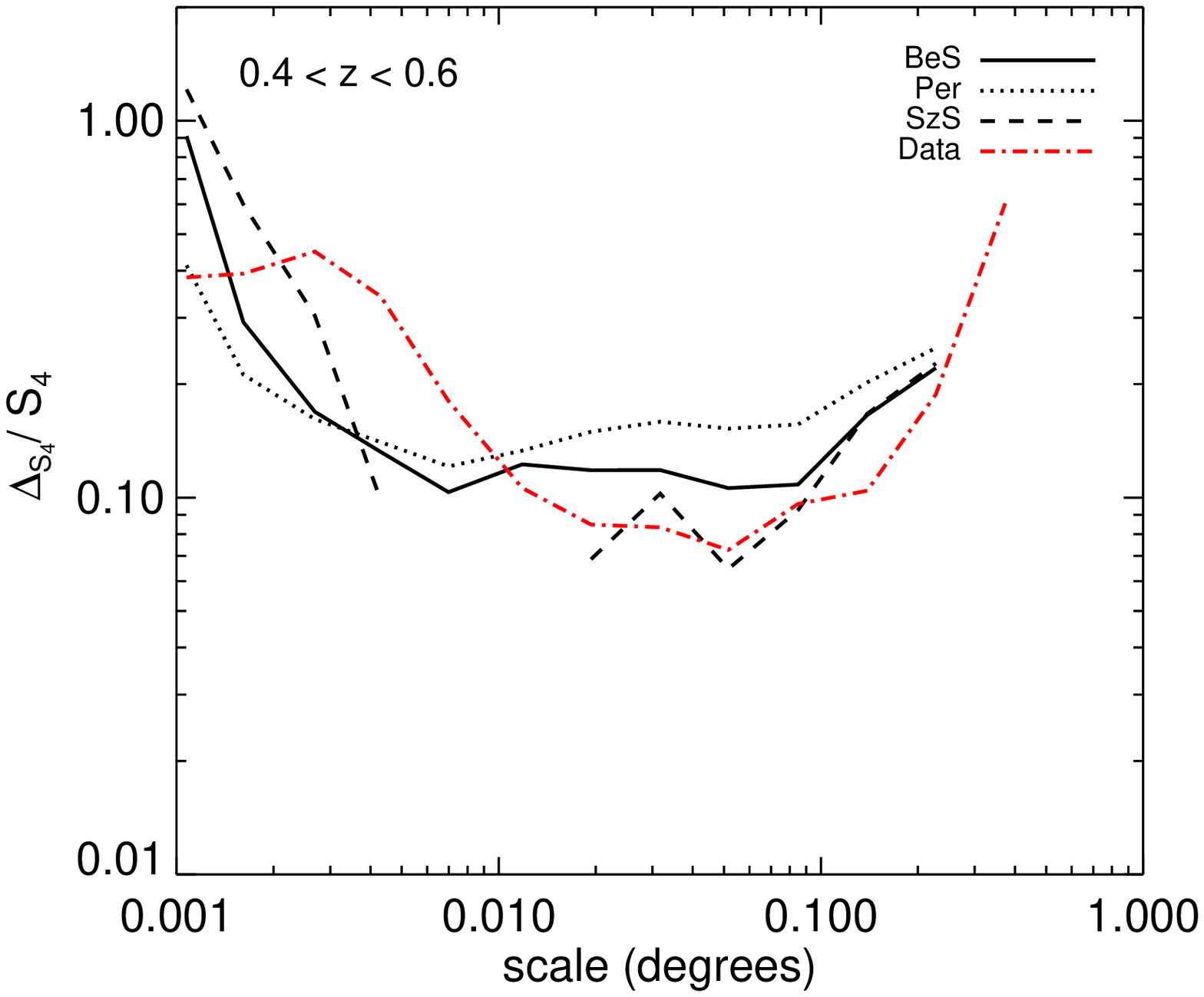}
      \end{tabular}
      \begin{tabular}{c@{}c@{}}
      \includegraphics[width=5.5cm]{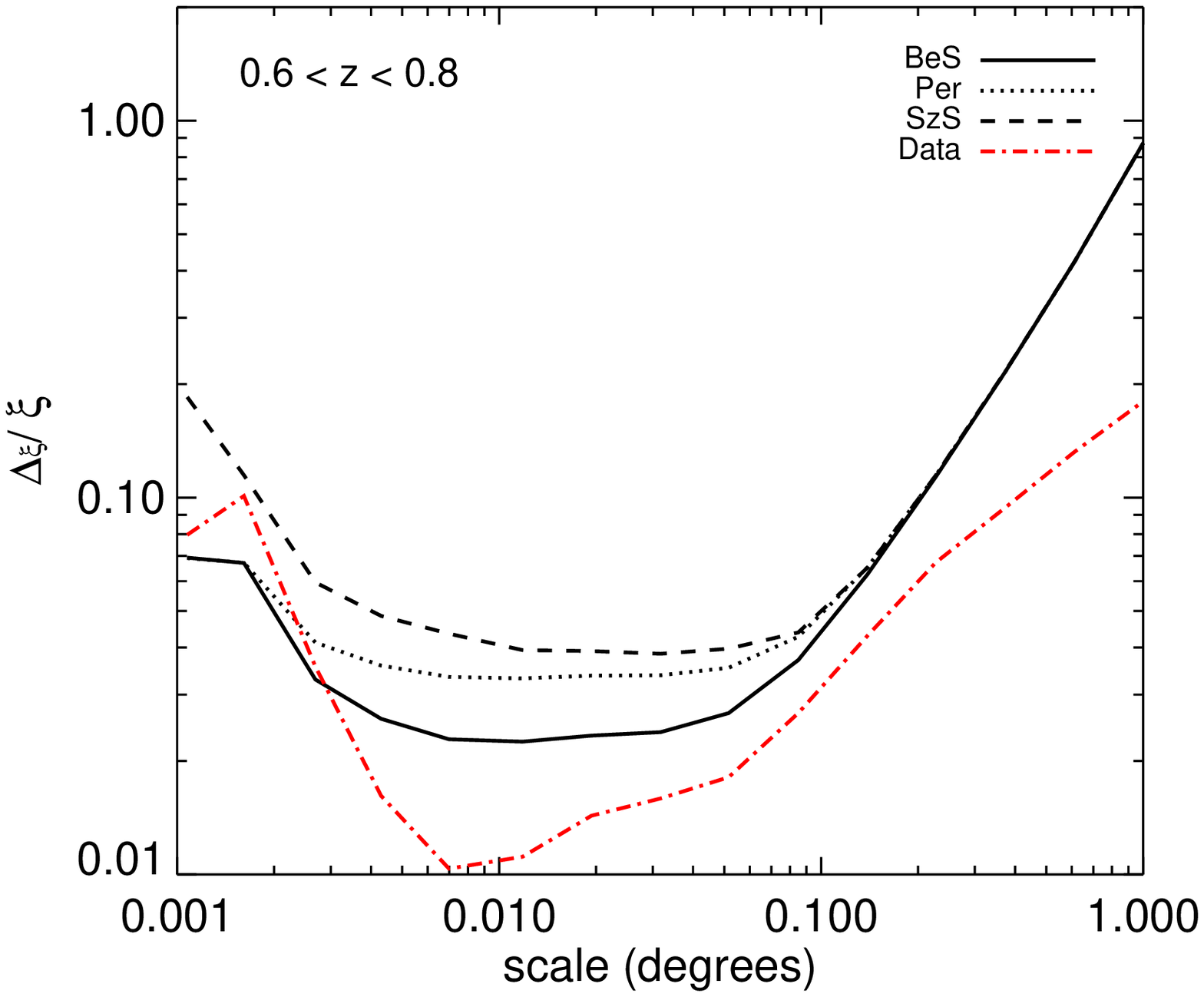}
      \includegraphics[width=5.5cm]{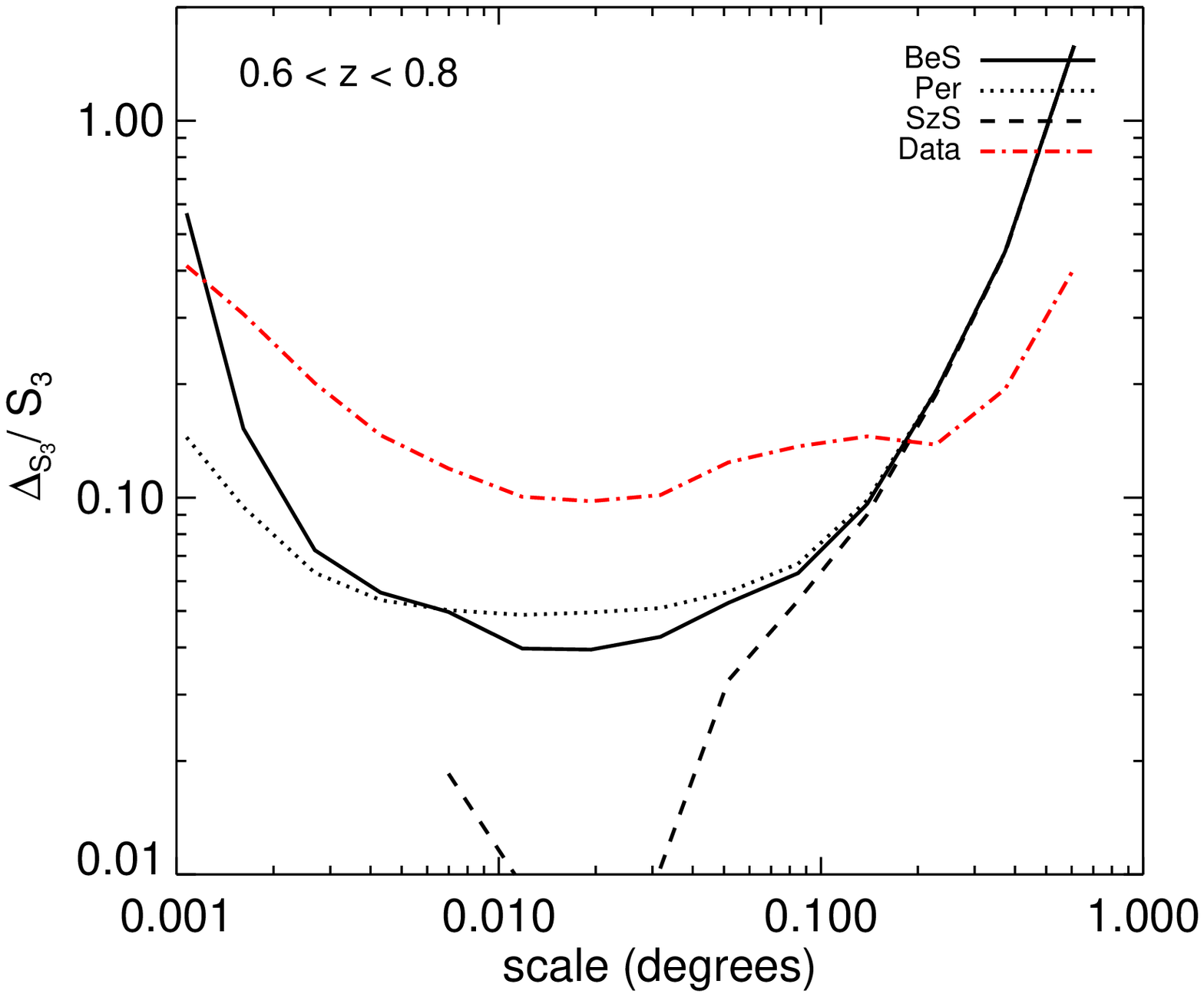}
      \includegraphics[width=5.5cm]{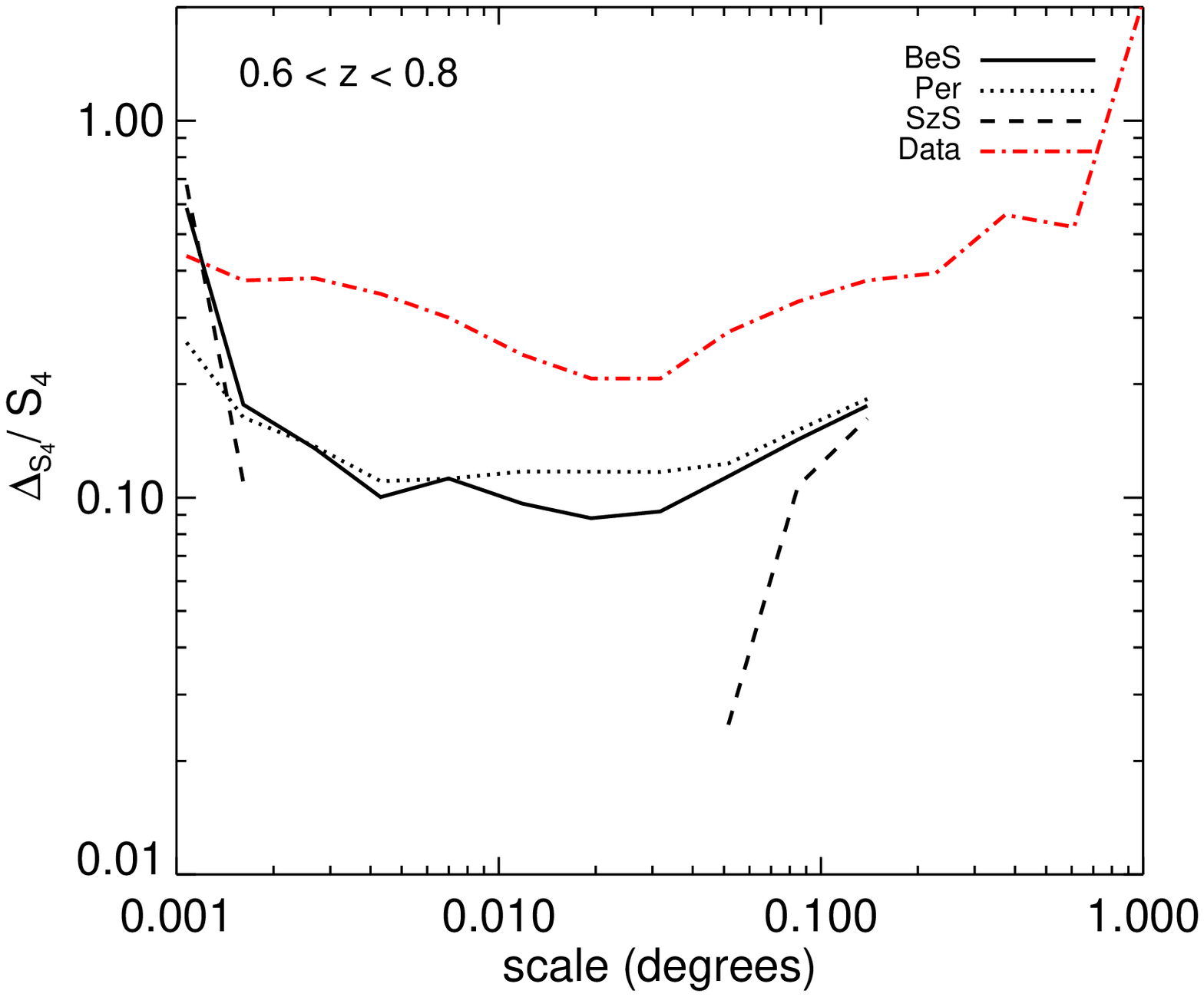}
      \end{tabular}
      \begin{tabular}{c@{}c@{}}
      \includegraphics[width=5.5cm]{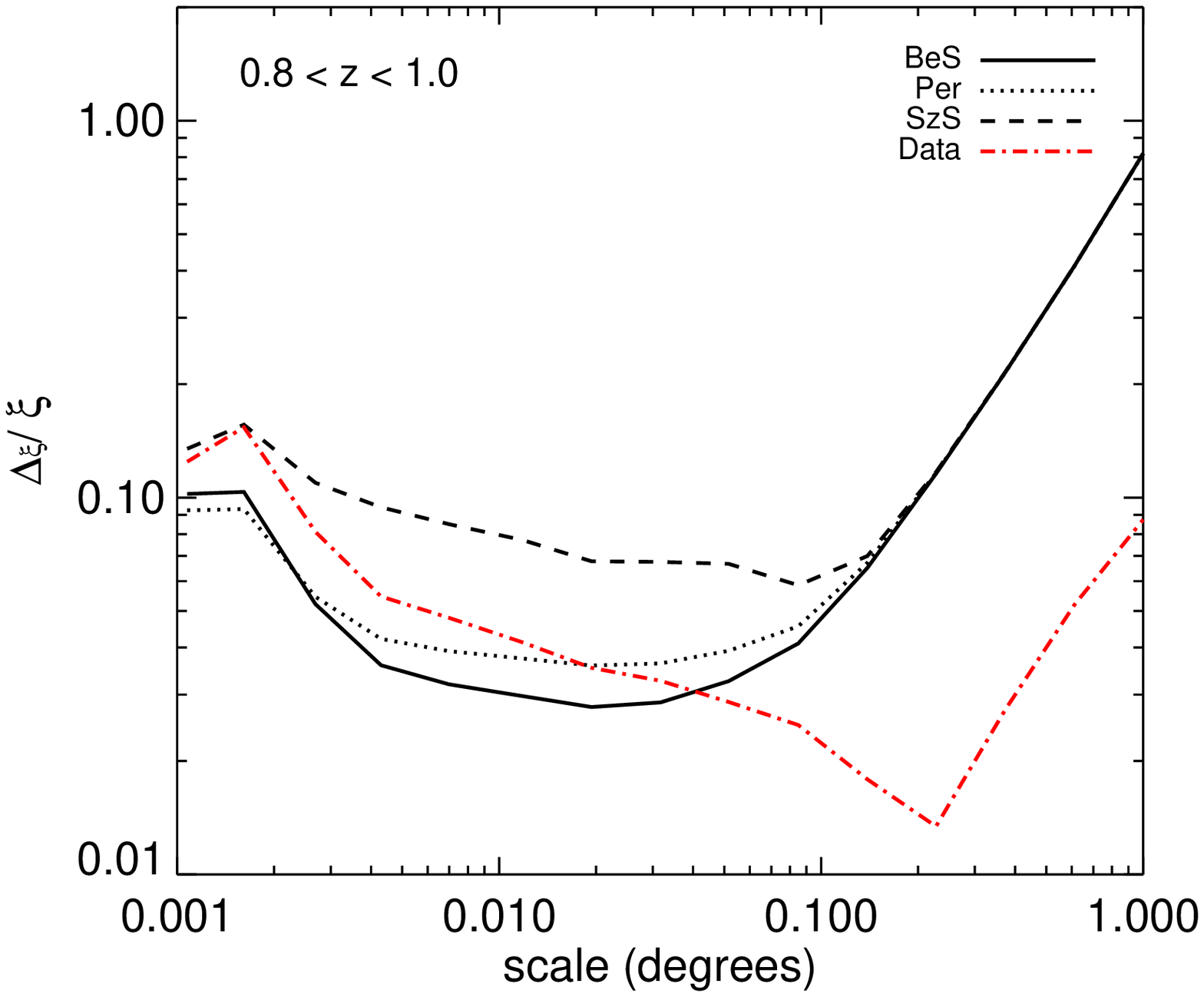}
      \includegraphics[width=5.5cm]{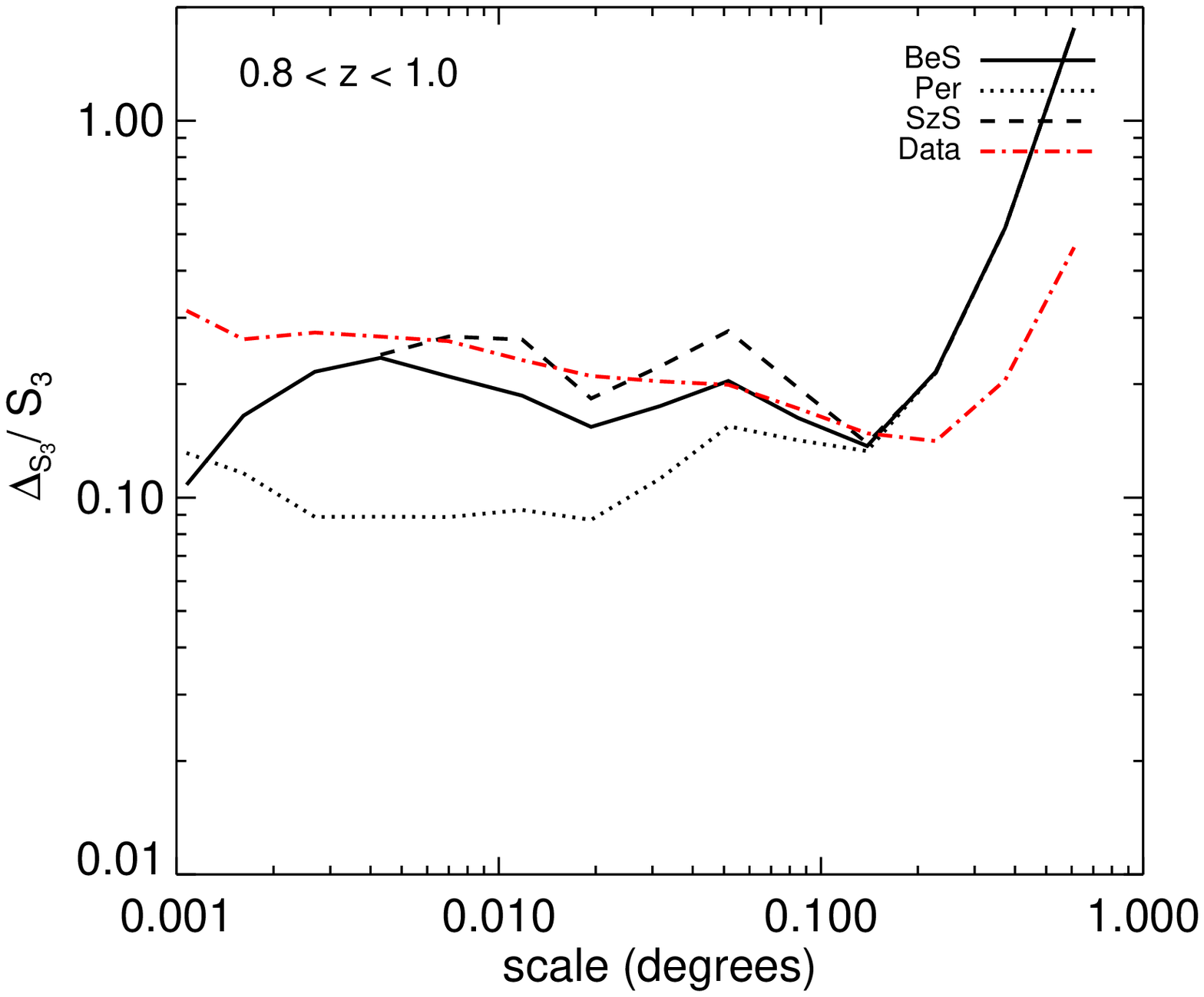}
      \includegraphics[width=5.5cm]{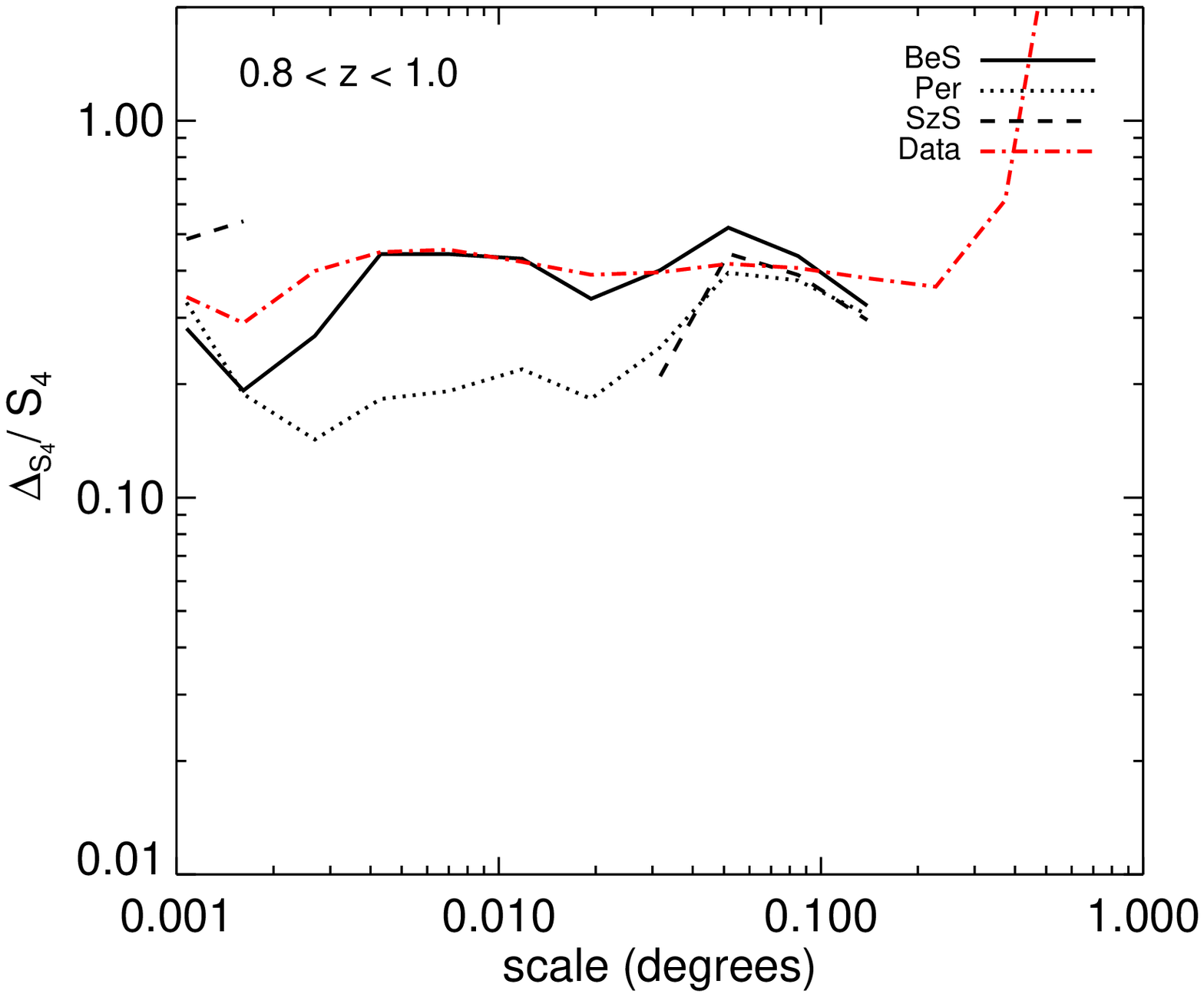}
    \end{tabular}
\end{center}

\caption{Comparison of theoretical errors to measured ones. The
  different panels from the left to the right represent the relative
  error on respectively the averaged two-point correlation function,
  $S_{3}$ and $S_{4}$ comparing the different models available in {\tt
    FORCE} (black curves) to the errors measured from the dispersion
  over the four fields as explained in Section
  \ref{sec:comb-heir-momem} (red curve).  For the theoretical
  predictions, three different models are available, SzS from
  \citet{Szapudietal2_93, Szapudietal93}, BeS from
  \citet{Bernardeauetal92} and Per uses perturbation theory
  predictions of \citet{Bernardeau94, Bernardeau96}. Each row of
  panels represents a different redshift bin going from the lowest one on
  the top to the highest one on the bottom. Some curves are
  interrupted where the models fail and give negative error bars.}
\label{figure:FORCE}
\end{figure*}

Figure~\ref{figure:FORCE} compares the theoretical errors on
$\bar{\xi}$, $S_{3}$ and $S_{4}$ derived with {\tt FORCE} to the
measured ones using the dispersion from the four fields. Three models
are considered, as detailed in the caption of the figure, to reflect
uncertainties on the theory. These theoretical predictions are
expected to be valid only when relative errors are small compared to
unity as consequence of the propagation error technique used to derive
it. Consequently, predictions at scales beyond $0.1$ degrees, where
edge effects start to dominate in the theory are expected to be less
accurate. With all these limitations in mind, one can see that the
agreement between theory and measurements is excellent except in the
redshift bin $0.6<z<0.8$ where the relative error seems to be slightly
underestimated/overestimated by the measurements when comparing to the
theory for $\bar{\xi}$ and the hierarchical moments respectively.

\begin{figure*}
  \begin{center}
    \begin{tabular}{c@{}c@{}}
      \includegraphics[width=6cm]{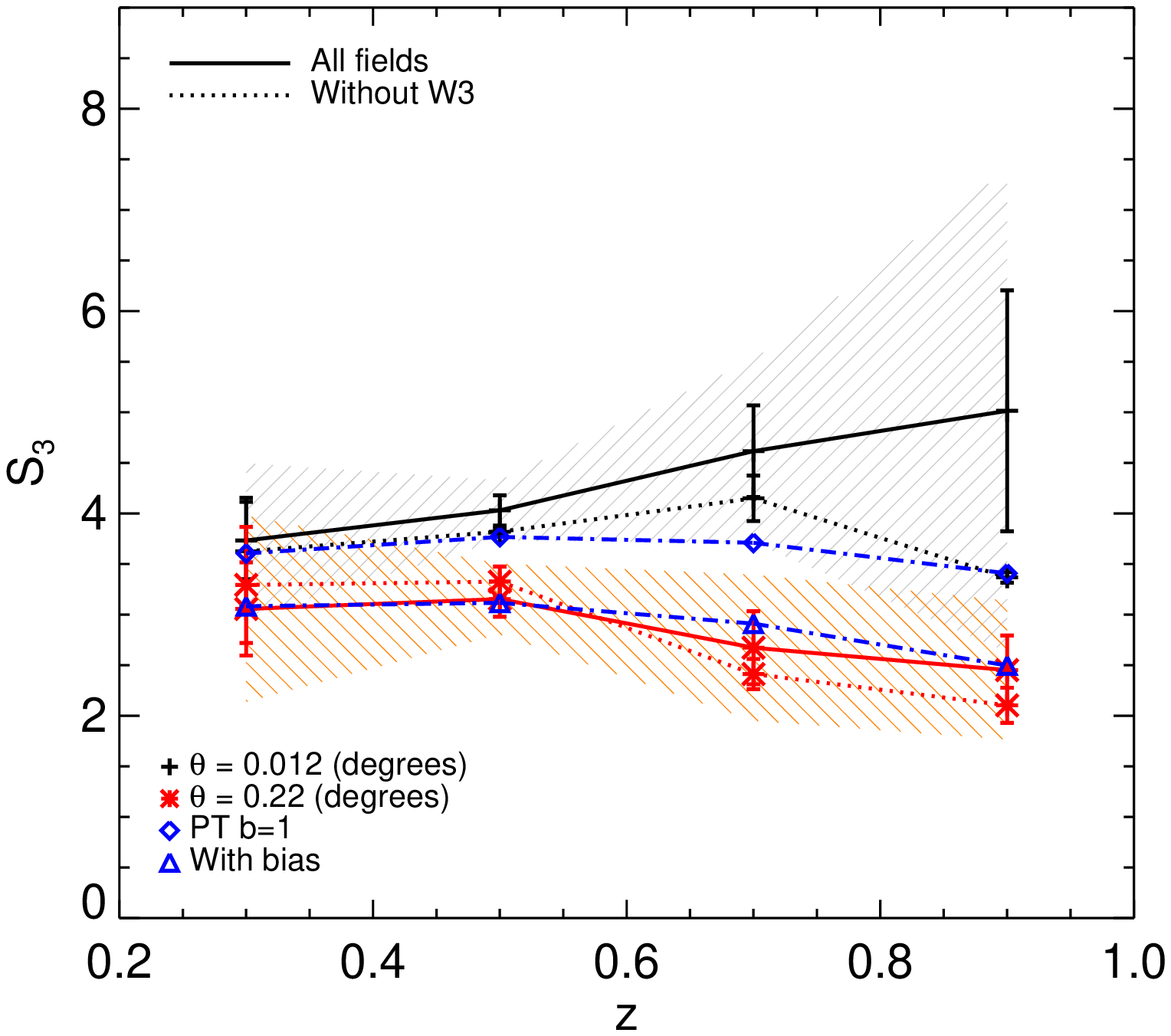}
      \includegraphics[width=6cm]{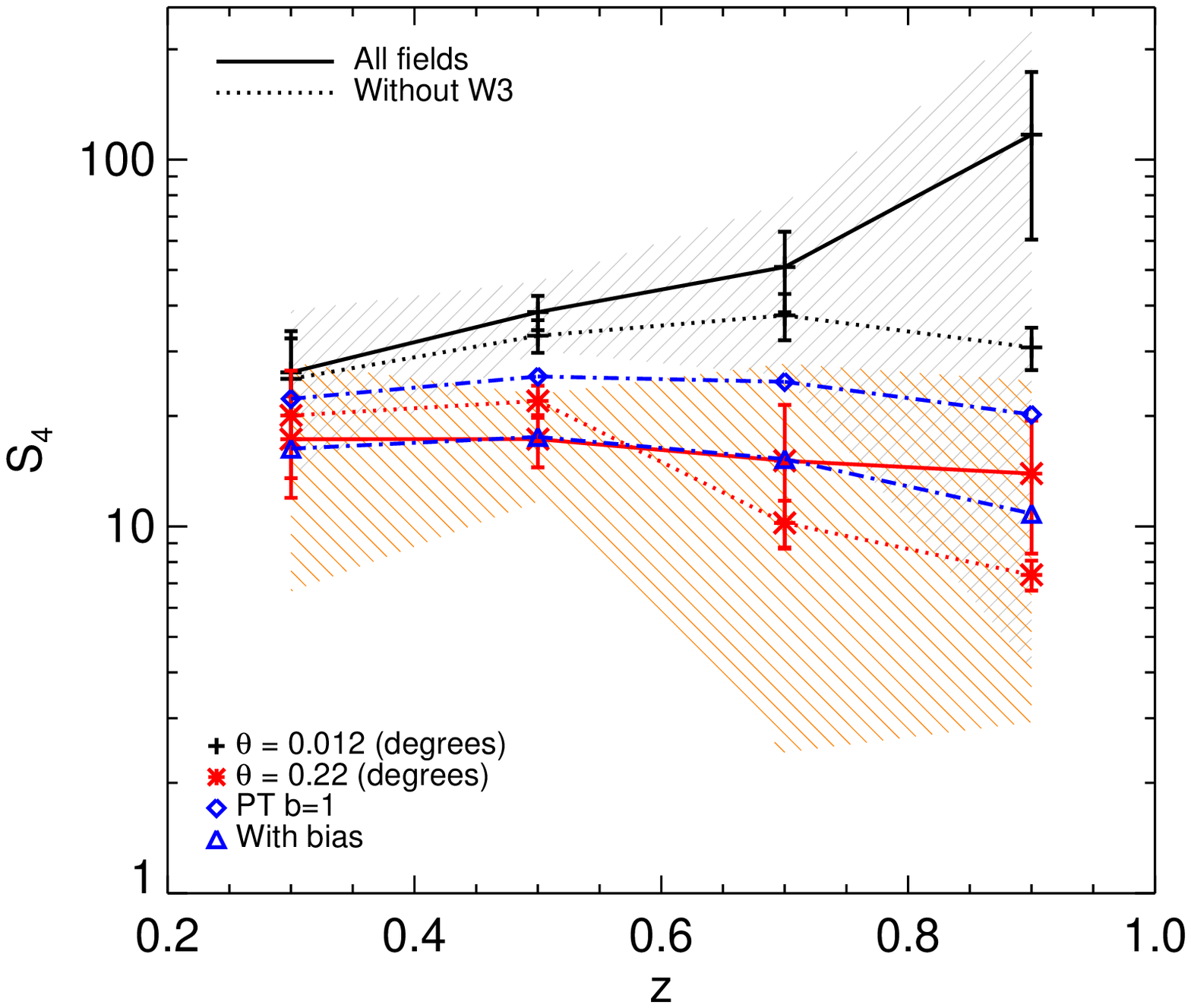}
      \includegraphics[width=6cm]{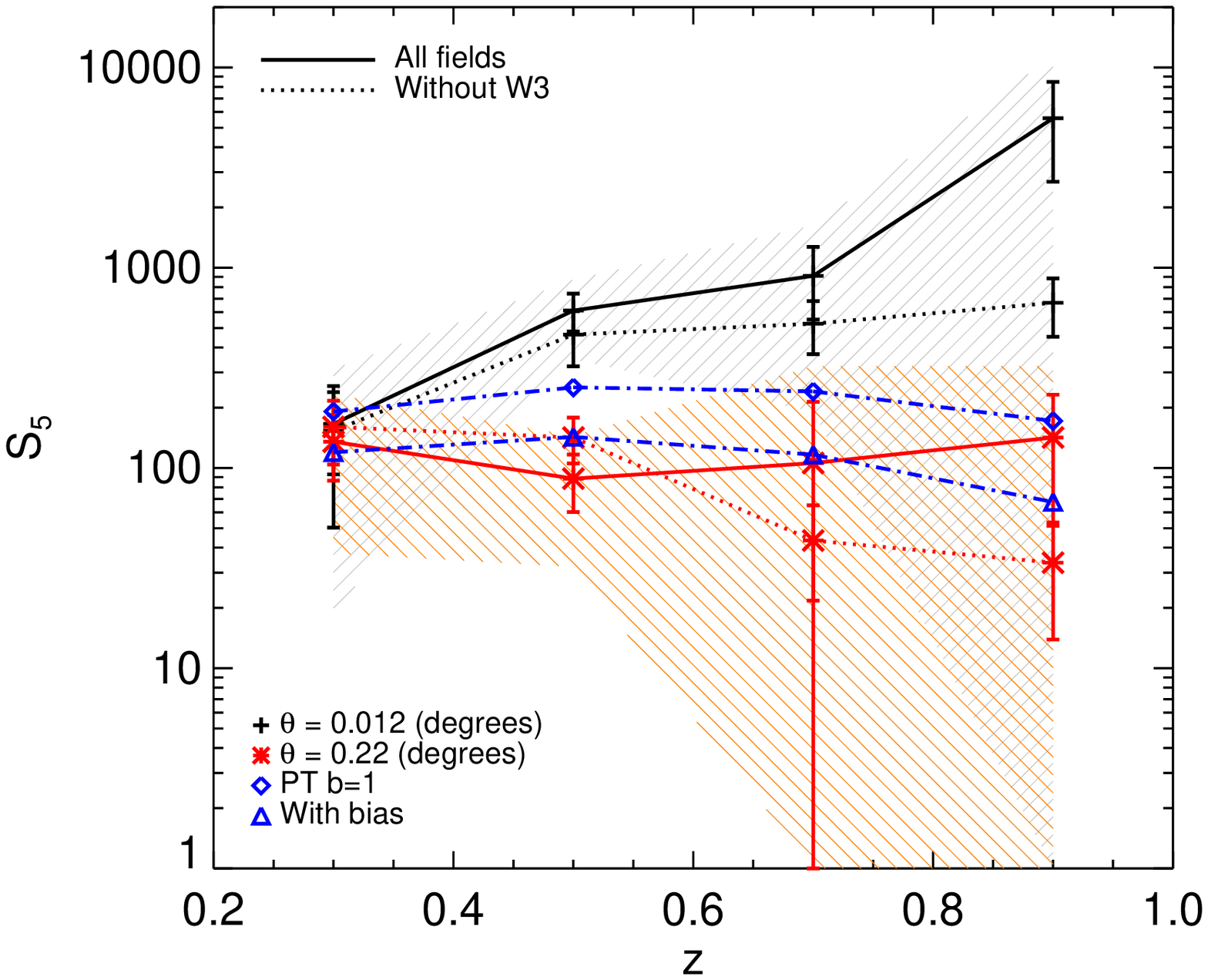}
      \end{tabular}
      \end{center}

      \caption{Evolution of $S_{3}$, $S_{4}$ and $S_{5}$ with redshift
        at non-linear and a weakly non-linear angular scales.  The
        black symbols linked with solid lines and those linked with
        dotted lines correspond to $\theta=0.012 \deg $ for
        W1+W2+W3+W4 and without W3 respectively.  The hashed grey
        region represent 2$\sigma$ around the solid black curves.
        Similarly for $\theta=0.22 \deg $ in red.  In addition,
        perturbation theory predictions for the scale $\theta=0.22 \deg $ are given in blue with linear
        bias (diamonds) and without (triangles).}
\label{figure:Snvsz}
\end{figure*}

The remarkable agreement between theoretical errors derived using the
standard ${\rm \Lambda}$CDM model and the measured weighted dispersion
between the four fields in the redshift bin $0.8<z<1$ demonstrates
that in fact W3 \emph{is not special}, from the statistical point of
view. This is confirmed by the examination of
Figure~\ref{figure:Snvsz}, which shows the hierarchical moments as
functions of redshift for two typical angular scales: $\theta=0.012
\deg $, which corresponds in practice to the non-linear regime, and
$\theta =0.22 \deg $, which is the largest possible reliable scale in
order to probe the weakly non linear regime. On this plot, one sees
that the $S_{n}$ measured after removing W3 are compatible within
2$\sigma$ with those obtained with all the fields.

\subsubsection{Evolution with redshift and comparison with
  perturbation theory}
\label{sec:zevolve}

\begin{figure*}
  \begin{center}
    \hbox{
      \includegraphics[width=8cm]{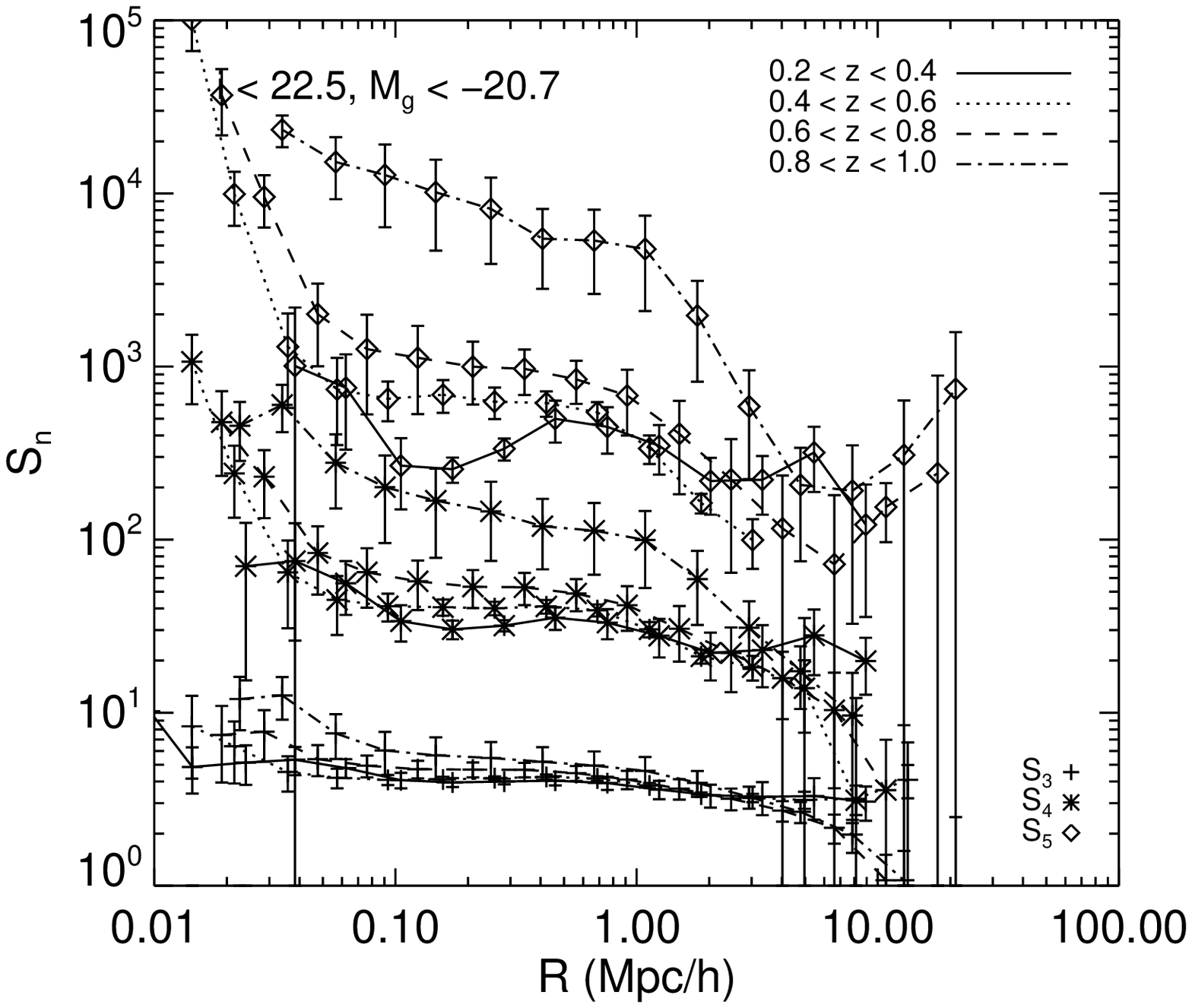}
      \includegraphics[width=8cm]{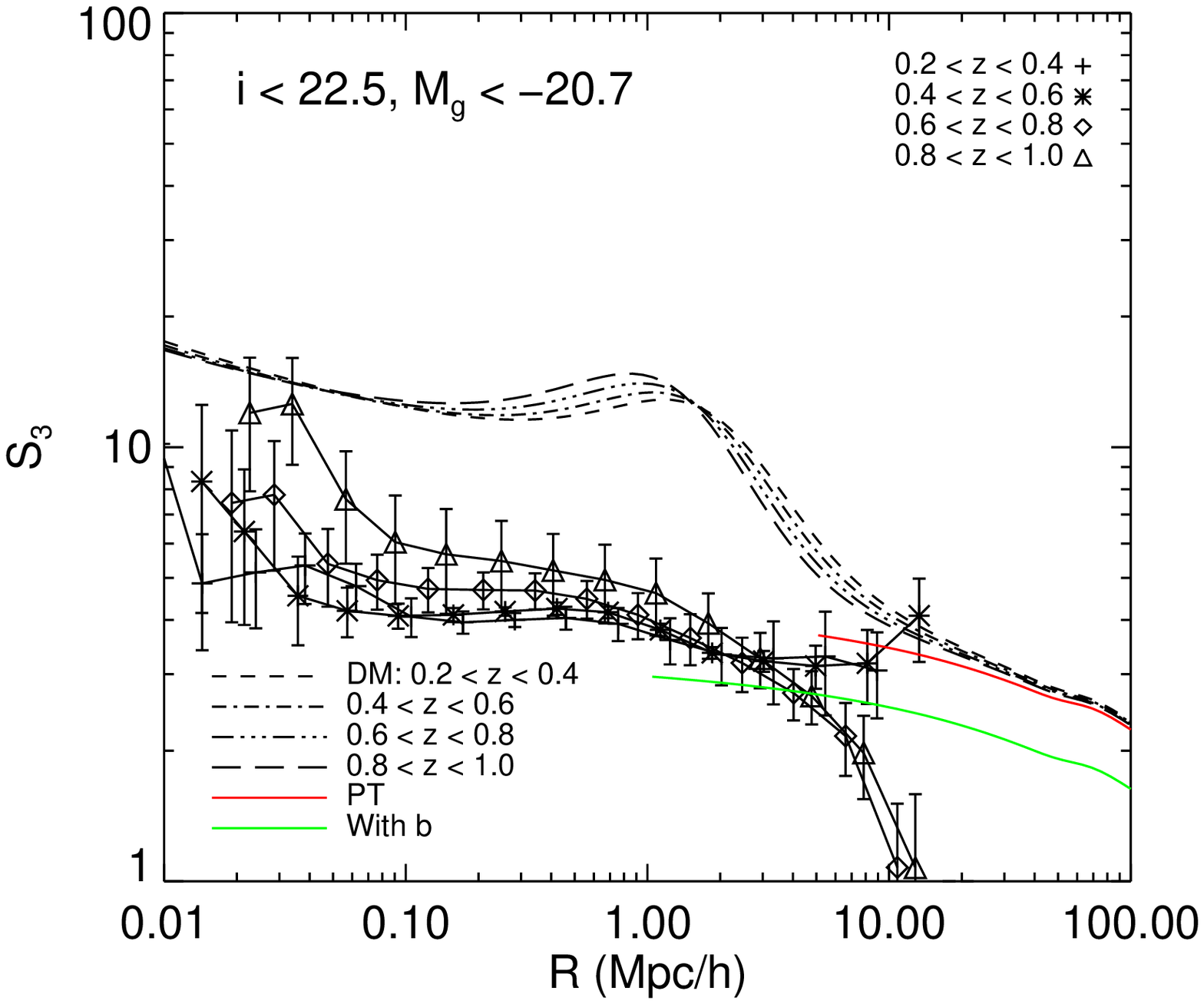}}
    \hbox{
      \includegraphics[width=8cm]{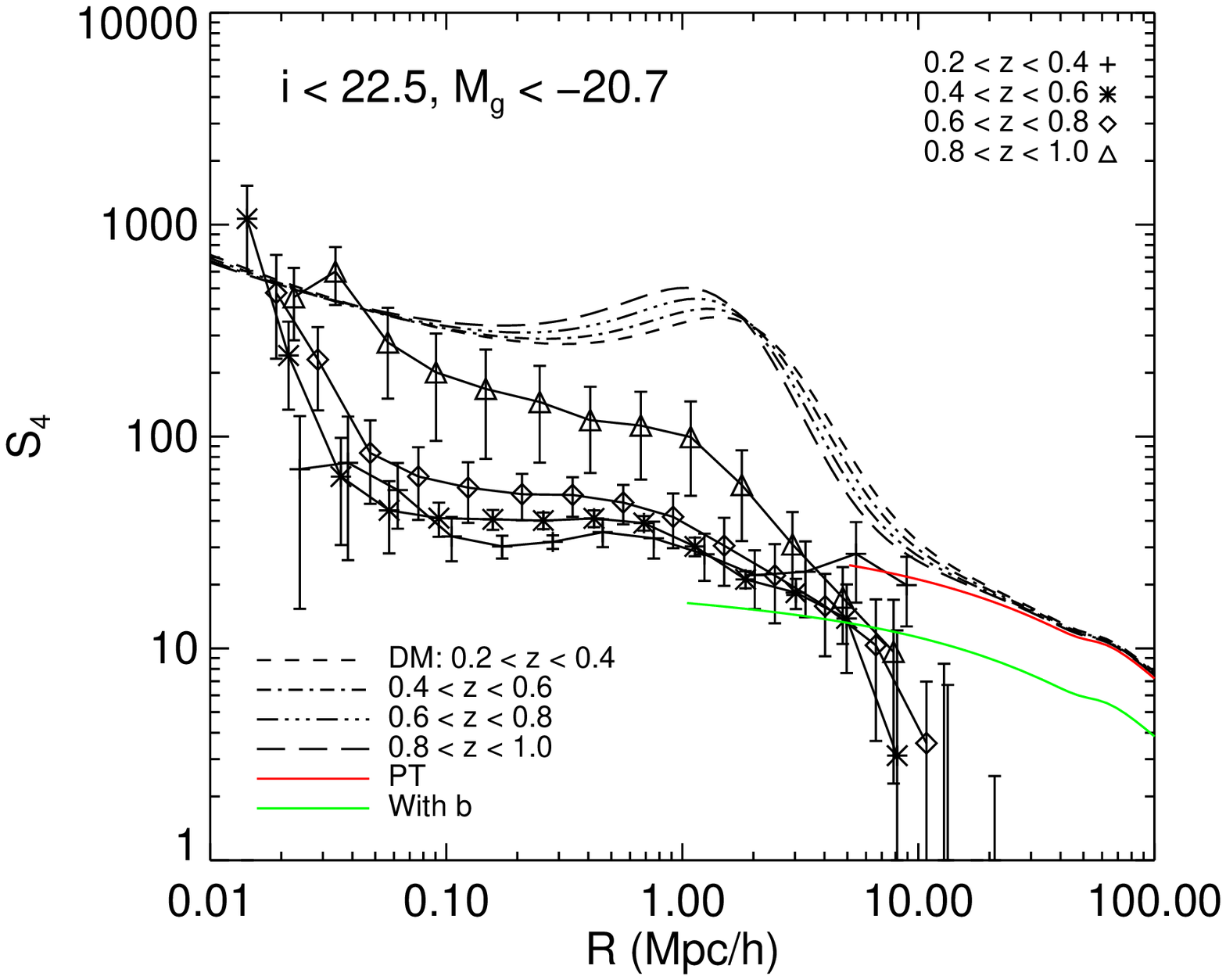}
      \includegraphics[width=8cm]{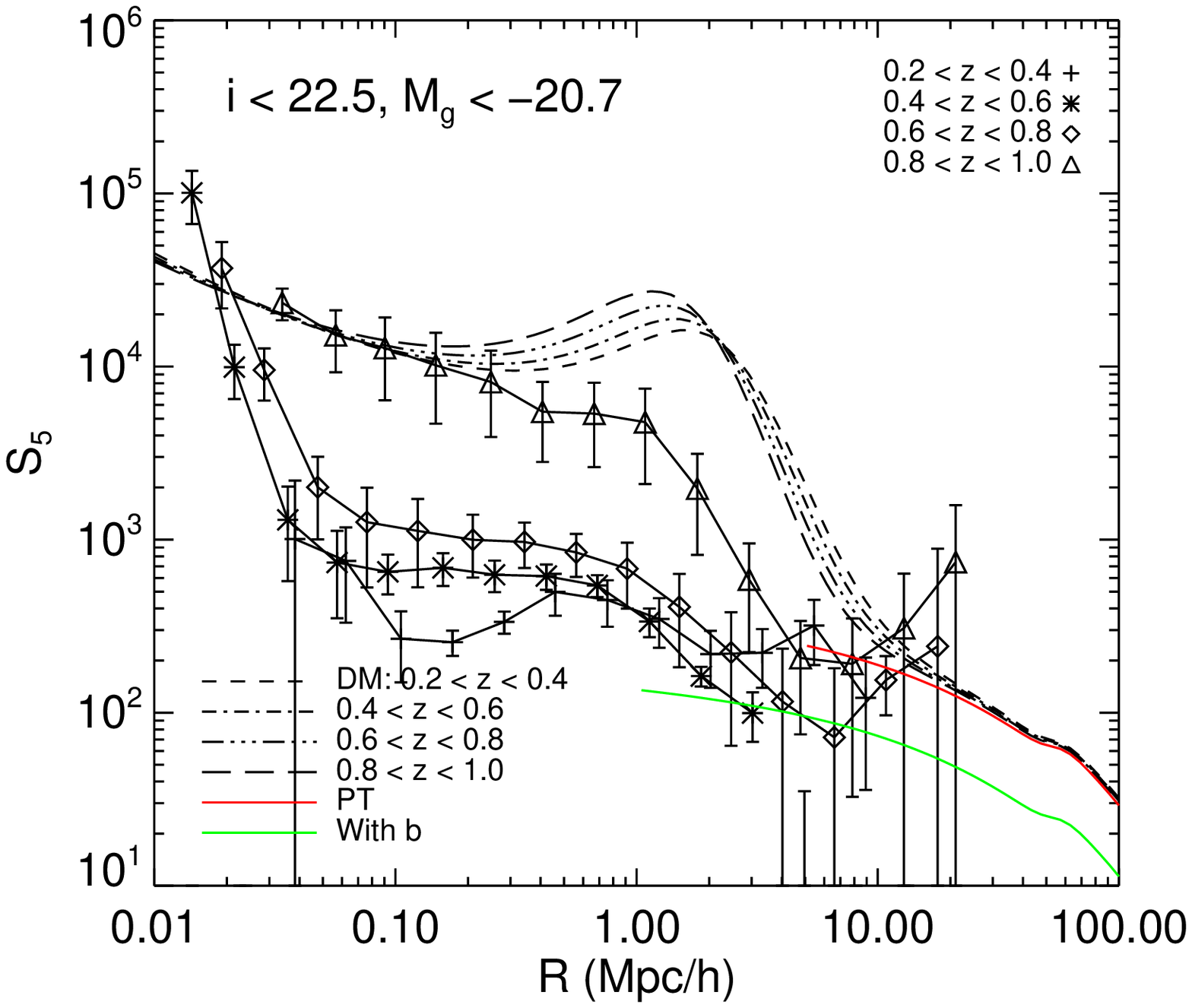}}
\end{center}
\caption{Three-dimensional $S_{n}$'s in the four redshift bins,
  together on the same plot (upper left panel) and represented
  individually (three remaining panels). The dashed and dot-dashed black curves on
  each panel gives dark matter halo model prediction. The coloured
  lines represent pure perturbation theory predictions using equations
  (\ref{eq:S3PT}), (\ref{eq:S4PT}) and (\ref{eq:S5PT}), in two cases:
  one without bias and one with the most extreme value of linear bias,
  $b=1.37$, found in the last redshift bin.}
\label{figure:SnsMpc}

\end{figure*} 

We have demonstrated that W3 does not represent a ``special event''
and that measurements with and without W3 are compatible within
2$\sigma$. This demonstrates that the redshift evolution trend in the
non-linear regime observed on left panel of Figure~\ref{figure:Sns}
and on Figure~\ref{figure:Snvsz} is indicative but inconclusive; this
statement holds when angular scales are converted to Mpc. From now on,
we do not consider furthermore W3 separately and present results only
for the full combination of the four fields.

Figure~\ref{figure:SnsMpc} shows the hierarchical moments in three
dimensional space as functions of physical scale after applying the
transformation described in Section~\ref{sec:projection}. The
measurements show the same behavior as before i.e. a plateau at small scales corresponding to the highly non-linear
  regime, and then tend to another lower plateau at large scales
  corresponding to the weakly non-linear regime. At scales
$\sim 10$ Mpc, we start to be sensitive to finite size
effects as we approach the size of the fields.

We now compare the measurements directly to perturbation theory
predictions \citep{Juszkiewicz93, Bernardeau94, Bernardeau94bis,
  Bernardeauetal04}:

\beqa
S_{3}^{\rm PT} = \frac{34}{7} + \gamma_{1},
\label{eq:S3PT}
\eeqa
\beqa
S_{4}^{\rm PT} = \frac{60712}{1323} + \frac{62}{3} \gamma_{1} + \frac{7}{3}
\gamma_{1}^{2} + \frac{2}{3} \gamma_{2},
\label{eq:S4PT}
\eeqa
\begin{eqnarray}
S_{5}^{\rm PT} &=& \frac{200575880}{305613} + \frac{1847200}{3969} \gamma_{1} +
\frac{6940}{63} \gamma_{1}^{2}  \nonumber \\
&&\quad{} +  \frac{235}{27} \gamma_{1}^{3} +
\frac{1490}{63} \gamma_{2} + \frac{50}{9} \gamma_{1} \gamma_{2} +
\frac{10}{27} \gamma_{3}
\label{eq:S5PT}
\end{eqnarray}
with $\gamma_{p} = {\rm d}^{p}\ln\sigma^{2}(R)/ {\rm d}^{p}\ln R$, where $\sigma^{2}(R)$ is the
linear variance of the 3D matter density field smoothed with a spherical
top-hat window of radius $R$.

The agreement between perturbation theory predictions and the
measurements at large scales is excellent and even better in a
striking way, as it can be seen on in Figures~\ref{figure:Snvsz} and
~\ref{figure:SnsMpc}, when simply taking into
account effects of linear bias:
\begin{equation}
S_{n,{\rm g}}=\frac{S_{n}}{b^{2(n-2)}},
\end{equation}

with the values of $b$ measured in Section~\ref{sec:comb-meas} and
given in Table~\ref{table:biasvalue}. With such an agreement well
within the error bars, we did not fit higher order bias terms
\citep{Fryetal93}; these are clearly consistent with zero at the level
of approximation of our measurements.

In addition, on Figure~\ref{figure:SnsMpc}, the full halo-model predictions for the
dark matter are given using the method of \cite{FCFBST11}. What is
interesting to notice here is that the halo model predictions and
measurements are consistent concerning the location of the transition
between the weakly non-linear regime and the highly non-linear regime.

\section{Conclusions}

We have measured the hierarchical moments of the galaxy distribution
$S_{3}$, $S_{4}$ and $ S_{5}$ in a volume-limited sample of more than
one million galaxies spanning the redshift range $z\sim1$ to
$z\sim2$. Our measurements were made in the latest release of the
CFHTLS wide and cover an effective area of $ 133 \, \textrm{deg}^2$.

We are able for the first time to make robust measurements of the
evolution of the hierarchical moments as functions of redshift within
a large range of angular scales. The survey consists of four
independent fields, allowing us to estimate the cosmic variance at
each bin and angular scale.

Our main results are as follows:

\begin{enumerate}
\item In the redshift bin $0.2<z<0.4$ for a sample with the same cut
  in absolute magnitude, our results are in excellent agreement with
  the SDSS demonstrating that our survey is a fair sample of the
  Universe;

\item This is confirmed by a comparison of the field to field scatter
  of the variance of counts in cells and the predicted hierarchical
  moments with the analytic ``cosmic errors'' predicted by
  \cite{Szapudi96, Szapudi99};

\item The hierarchical moments that we measure show two regimes
  separated by a transition region: first, a plateau at small scales
  ($ \la 1 \, h^{-1} \Mpc$) corresponding to the highly non-linear
  regime, and then the data tend to another lower plateau at large scales ($ \ga 7
  h^{-1} \Mpc$) corresponding to the weakly non-linear regime;

\item At non-linear scales, the amplitude of the hierarchical moments
  increase with redshift: however, a more accurate error analysis
  demonstrates that our results are also consistent with no redshift
  evolution within 2$\sigma$;

\item In the weakly non-linear regime measurements are fully
  consistent with the perturbation theory predictions when a linear
  bias term (measured from our data) is taken into
  account. Higher-order bias terms corrections \citep{Fryetal93} are
  negligible compared to linear bias given the precision of our
  measurements; 

\item The position of the transition between the non-linear and the
  weakly non-linear regime $1 \la R \la 10 \, h^{-1} \Mpc $ is fully
  compatible with halo model predictions.
\end{enumerate}

A robust conclusion concerning the redshift evolution of the
hierarchical moments in the non-linear regime, and the corresponding
implications for galaxy formation and evolution, is challenging.  This
is because the richest clusters in our sample can dramatically
influence higher order measurements. A complete understanding of this
effect will require a detailed modelling of the counts in cells
probability using the halo model formalism and taking into account the
halo mass of the most massive structure in our survey. This will be
the subject of a subsequent work.

\section{Acknowledgments}

We would like to thank Ashely Ross for providing his code to calculate
the counts in cells for SDSS DR7 and the associated catalogs and
advice.  We acknowledge the TERAPIX team and Raphael Sadoun for his
useful help and suggestions. H. J. McCracken acknowledges support from the `Programme national
cosmologie et galaxies''.  JF thanks the IAP for hospitality during
his work. H. J. McCracken and MW acknowledge the use of TERAPIX
computing resources.

%\clearpage
\bibliographystyle{mn2e}
\bibliography{snpaper,papers}

\appendix

\section{Weighted statistics}
\label{app:stat}

In this Appendix, we describe the way we estimate errors of the
combined quantities.

Let $x_{i}$, $i=1,\dots,n$, be independent variables, 
$\lexp x_{i}x_{j} \rexp = \lexp x_{i} \rexp \lexp x_{j} \rexp$ 
for $i \neq j$,
with a common mean $\lexp x_{i} \rexp = \mu$ and variances $\lexp
(x_{i} - \mu)^{2} \rexp = \sigma_{i}^{2}$. 
Conceptually, the $x_i$ are observations of some quantity 
in independent fields.
Let $w_{i}$ be a set of weights we define 
and use to compute a weighted mean 
\begin{equation}
\xbar \equiv \frac{1}{w_{\rm tot}} \sum_{i=1}^{n} w_{i}x_{i}, 
\label{xbar}
\end{equation}
and dispersion 
\begin{equation}
 \sum_{i=1}^{n} w_{i}(x_{i}-\xbar)^{2},
\end{equation}
with $ w_{\rm tot} \equiv \sum w_{i} $.
We seek a relation between the dispersion among the samples, 
$ \Delta x $, and the uncertainty in the mean, $ \Delta \xbar $.
The expectation of the weighted mean is 
\beq
\lexp \xbar \rexp= \frac{1}{w_{\rm tot}} \sum_{i=1}^{n}
w_{i} \lexp x_{i} \rexp = \mu , 
\eeq
and the expectation of the weighted variance is 
\beq
\lexp (\Delta x)^{2} \rexp= \frac{1}{w_{\rm tot}} \sum_{i=1}^{n}
w_{i} \lexp x_{i}^{2} -2\xbar x_{i} + \xbar^{2} \rexp .
\eeq
The terms in the sum can be expressed in terms of $\mu$ 
and the $ \sigma_i^2 $, 
\beq
\lexp x_{i}^{2} \rexp = \mu^{2} + \sigma_{i}^{2} , 
\eeq
\beqa
\lexp \xbar x_{i} \rexp &=& \frac{1}{w_{\rm tot}}
\sum_{j} w_{j} \lexp x_{i}x_{j} \rexp \nonumber \\
&=& \frac{1}{w_{\rm tot}} \Bigl[w_{i}(\mu^{2} + \sigma_{i}^{2}) + \sum_{i
  \neq j} w_{j}\mu^{2}\Bigr] 
\mu^{2} + \frac{w_{i}\sigma_{i}^{2}}{w_{\rm tot}} , 
\eeqa
and 
\beqa
\lexp \xbar^{2} \rexp &=& \frac{1}{w_{\rm tot}^{2}}
\sum_{jk} w_{j}w_{k}\lexp x_{j}x_{k} \rexp \nonumber \\
&=& \frac{1}{w_{\rm tot}^{2}} \Bigl[\sum_{j} w_{j}^{2}(\mu^{2} +
\sigma_{j}^{2}) + \sum_{j \neq k} w_{j}w_{k}\mu^{2} \Bigr] \nonumber \\
&=&\mu^{2} + \frac{1}{w_{\rm tot}^{2}} 
{\sum w_{j}^{2}\sigma_{j}^{2}} .
\eeqa
Adding these together, the terms in $\mu$ cancel, leaving 
\beqa
\lexp (\Delta x)^{2} \rexp &=& \frac{1}{w_{\rm tot}} \sum_{i} w_{i} 
\Bigl[\sigma_{i}^{2} - \frac{2w_{i}\sigma_{i}^{2}}{w_{\rm tot}} + 
\frac{\sum_{j} w_{j}^{2}\sigma_{j}^{2}}{w_{\rm tot}^{2}}\Bigr] 
\nonumber \\
&=& \frac{1}{w_{\rm tot}} {\sum_{i} w_{i}\sigma_{i}^{2}}
- \frac{1}{w_{\rm tot}^{2}}{\sum_{i} w_{i}^{2}\sigma_{i}^{2}} .
\eeqa
The variance in $\xbar$ is similarly computed to be 
\beq
\lexp (\Delta \xbar)^{2} \rexp 
= \lexp \xbar^{2} \rexp - \lexp \xbar \rexp^{2} 
= \frac{1}{w_{\rm tot}^{2}} \sum_{i} w_{i}^{2}\sigma_{i}^{2} .
\eeq
In several cases this leads to a simple relation 
between $ \Delta x $ and $ \Delta \xbar $.
If all the $w_{i}$ are equal then 
\beq
\lexp (\Delta x)^{2} \rexp = \frac{n-1}{n^2} \sum_{i}\sigma_{i}^{2}, 
\eeq
and 
\beq
\lexp (\Delta \xbar)^{2} \rexp= \frac{1}{n^2} \sum_{i}\sigma_{i}^{2} .
\eeq
For $w_{i}= 1/ \sigma_{i}^{2}$
\beq
\lexp (\Delta x)^{2} \rexp = \frac{n-1}{\sum_{i} 1/ \sigma_{i}^{2}}
\eeq
and 
\beq
\lexp (\Delta \xbar)^{2} \rexp= \frac{1}{\sum_{i} 1/ \sigma_{i}^{2}}.
\eeq
The weights $w_{i}= 1/ \sigma_{i}^{2}$ are the minimum variance
estimator, and choosing the effective area as weights is the 
best approximation to this.\footnote{at least in the Poisson approximation, 
when $x_{i}=F_{1}$ measured in field $i$}
Therefore, we use \eq{xbar} with weights $ w_i = S_{\eff,i} $ 
as the average over fields, and adopt as our estimate of uncertainty 
\beqa
\lexp (\Delta \xbar)^{2} \rexp= \frac{\lexp (\Delta x)^{2} \rexp}{n-1} .
\eeqa

\section{$S_{n}$ measurements}

\begin{table*}
\begin{minipage}{195mm}
\centering
\begin{tabular}{ccccccc}
$0.2 < z < 0.4$ & & & & & & \\
\hline
$\theta $ & $S_{3}$ & $ \sigma_{S_{3}}$ & $ S_{4}$ & 
 $\sigma_{S_{4}}$ & $S_{5}$ & $\sigma_{S_{5}}$ \\
\hline
    0.0011 &         9.34 &         4.84 &        $-8.92$  &         3.96 &       $-15.5$ &         7.99 \\
    0.0016 &         4.45 &         1.33 &        $-4.29$  &         0.80 &        $-6.8$ &         1.85 \\
    0.0027 &         4.72 &         1.21 &        54.7    &        42.7 &       $-84.0$ &        44.7 \\
    0.0043 &         4.90 &         0.91 &        58.6    &        38.3 &       625 &       735 \\
    0.0070 &         4.39 &         0.55 &        43.6 &        15.0 &       468 &       262 \\
    0.0118 &         3.73 &         0.38 &        26.3 &         6.22 &       166 &        73.1 \\
    0.0193 &         3.62 &         0.21 &        23.6 &         2.91 &       158 &        26.4 \\
    0.0317 &         3.67 &         0.14 &        24.8 &         1.83 &       208 &        30.3 \\
    0.0516 &         3.71 &         0.22 &        27.6 &         4.12 &       310 &        84.2 \\
    0.0849 &         3.59 &         0.31 &        25.8 &         5.11 &       278 &        83.2 \\
    0.1392 &         3.30 &         0.40 &        21.7 &         5.44 &       216 &        68.8 \\
    0.2273 &         3.06 &         0.46 &        17.3 &         5.33 &       135 &        48.9 \\
    0.3729 &         2.99 &         0.65 &        17.9 &         7.01 &       138 &        51.2 \\
    0.6105 &         3.02 &         0.81 &        21.8 &         8.96 &       198 &        80.7 \\
    1.0001 &         2.81 &         0.63 &        15.5 &         5.60 &        75.7 &        53.5 \\
\hline
\end{tabular}
\begin{tabular}{ccccccc}
$0.4 < z < 0.6$ & & & & & & \\
\hline
$\theta $ & $S_{3}$ & $ \sigma_{S_{3}}$ & $ S_{4}$ & 
 $\sigma_{S_{4}}$ & $S_{5}$ & $\sigma_{S_{5}}$ \\
\hline
     0.0011 &         8.18 &         4.10 &      1007 &       435 &     89701 &     30529 \\
      0.0016 &         6.27 &         2.45 &       228 &       101 &      8823 &      3051 \\
      0.0027 &         4.46 &         1.03 &        61.1 &        32.1 &      1157 &       645 \\
      0.0043 &         4.12 &         0.54 &        42.3 &        15.8 &       656 &       343 \\
      0.0070 &         4.01 &         0.27 &        38.9 &         7.14 &       579 &       150 \\
      0.0118 &         4.03 &         0.15 &        38.4 &         4.11 &       611 &       131 \\
      0.0193 &         4.10 &         0.09 &        37.9 &         3.40 &       557 &       115 \\
      0.0317 &         4.17 &         0.14 &        38.8 &         3.50 &       547 &        91.4 \\
      0.0516 &         4.09 &         0.13 &        36.9 &         2.93 &       484 &        70.3 \\
      0.0849 &         3.72 &         0.12 &        28.6 &         2.89 &       299 &        56.4 \\
      0.1392 &         3.30 &         0.10 &        20.0 &         1.86 &       145 &        18.8 \\
      0.2273 &         3.15 &         0.17 &        17.3 &         2.81 &        88.5 &        28.2 \\
      0.3729 &         3.07 &         0.35 &        13.1 &         5.85 &       $-10.1$ &        41.4 \\
      0.6105 &         3.12 &         0.60 &         2.94 &         8.54 &      $-249$ &        43.0 \\
      1.0001 &         4.01 &         0.87 &        $-5.00$ &        11.4 &      $-595$ &       241 \\  
\hline
\end{tabular}
\vskip 0.5cm
\begin{tabular}{ccccccc}
$0.6 < z < 0.8$ & & & & & & \\
\hline
$\theta $ & $S_{3}$ & $ \sigma_{S_{3}}$ & $ S_{4}$ & 
 $\sigma_{S_{4}}$ & $S_{5}$ & $\sigma_{S_{5}}$ \\
\hline
      0.0011 &         7.32 &         3.43 &       455 &       232 &     33737 &     13985 \\
      0.0016 &         7.64 &         2.53 &       220 &        93.4 &      8710 &      2910 \\
      0.0027 &         5.30 &         1.08 &        79.8 &        33.9 &      1832 &       916 \\
      0.0043 &         4.86 &         0.70 &        61.8 &        23.2 &      1151 &       668 \\
      0.0070 &         4.64 &         0.54 &        54.7 &        17.4 &      1025 &       541 \\
      0.0118 &         4.62 &         0.45 &        51.0 &        12.6 &       910 &       360 \\
      0.0193 &         4.60 &         0.44 &        50.6 &        10.7 &       885 &       260 \\
      0.0317 &         4.40 &         0.44 &        46.6 &         9.81 &       769 &       214 \\
      0.0516 &         4.05 &         0.49 &        40.0 &        11.5 &       620 &       255 \\
      0.0849 &         3.58 &         0.48 &        29.1 &        10.3 &       373 &       206 \\
      0.1392 &         3.13 &         0.45 &        21.1 &         8.56 &       203 &       144 \\
      0.2273 &         2.67 &         0.36 &        15.1 &         6.34 &       106 &       108 \\
      0.3729 &         2.13 &         0.41 &         9.87 &         6.38 &        65.7 &        99 \\
      0.6105 &         1.06 &         0.43 &         3.41 &         3.25 &       141 &        53 \\
      1.0001 &        $-0.69$ &    0.33 &        $-5.60$  &      2.64 &       221 &       588 \\
\hline
\end{tabular}
\begin{tabular}{ccccccc}
$0.8 < z < 1.0$ & & & & & & \\
\hline
$\theta $ & $S_{3}$ & $ \sigma_{S_{3}}$ & $ S_{4}$ & 
 $\sigma_{S_{4}}$ & $S_{5}$ & $\sigma_{S_{5}}$ \\
\hline
      0.0011 &        11.0 &         3.75 &       365 &       136 &      $-819$ &       226 \\
      0.0016 &        11.5 &         3.18 &       483 &       147 &     16011 &      3325 \\
      0.0027 &         6.96 &         2.01 &       223 &       102 &     10415 &      4062 \\
      0.0043 &         5.54 &         1.55 &       161 &        84.7 &      8770 &      4397 \\
      0.0070 &         5.20 &         1.41 &       135 &        71.8 &      6970 &      3767 \\
      0.0118 &         5.01 &         1.19 &       117 &        56.4 &      5579 &      2893 \\
      0.0193 &         4.77 &         1.02 &        96.0 &        42.0 &      3757 &      1829 \\
      0.0317 &         4.54 &         0.93 &        90.5 &        40.2 &      3664 &      1865 \\
      0.0516 &         4.23 &         0.85 &        79.9 &        37.7 &      3271 &      1836 \\
      0.0849 &         3.60 &         0.62 &        47.5 &        21.7 &      1352 &       792 \\
      0.1392 &         2.99 &         0.44 &        24.9 &        10.5 &       403 &       249 \\
      0.2273 &         2.45 &         0.34 &        13.9 &         5.50 &       142 &        90.5 \\
      0.3729 &         1.81 &         0.39 &         7.74 &         5.89 &       131 &       109 \\
      0.6105 &         0.99 &         0.47 &        $-1.92$ &    8.71 &       211 &       225 \\
      1.0001 &        $-0.36$ &    0.95 &       $-15.02$ &   17.0 &       508 &       575 \\
\hline
\end{tabular}
\end{minipage}
\caption{
Measured $S_n$ and their error bars in the bins $0.2<z<0.4$,
$0.4<z<0.6$, $0.6<z<0.8$ and $0.8<z<1$, respectively on the
upper left, upper right, bottom left and bottom right table. On each
table, the first
column refers to the angular size of the square cells used to perform the
measurements. The next columns display successively $S_{3}$ and its error
bars, $S_{4}$ and its error bars and $S_{5}$ and its error bars. The
counts in cells
method used to perform the measurements is described in
\S~\ref{sec:BMW}. The way the four fields are combined and the
corresponding error bars are computed is detailed in
\ref{sec:comb-heir-momem}.}
\label{table:Sns}
\end{table*}

\end{document}